\documentclass[twocolumn,preprintnumbers,nofootinbib]{revtex4}
\usepackage{graphicx}
\usepackage{amssymb}
\usepackage{color}
\usepackage{enumerate}
\def\beq{\begin{equation}}
\def\eeq{\end{equation}}
\def\beqn{\begin{eqnarray}}
\def\eeqn{\end{eqnarray}}

\renewcommand{\texttt}{{}}
\newcommand{\be}{\begin{eqnarray}}
\newcommand{\ee}{\end{eqnarray}}

\begin{document}

%\title{The microstructure of a quantum spacetime}
%\title{Complete Nonlocal Quantum Gravity}
\title{%Multidimensional 
Super-renormalizable Multidimensional Quantum 
Gravity}
\author{Leonardo Modesto}
%\email{lmodesto@perimeterinstitute.ca}
%\thanks{Electronic address: lmodesto@perimeterinstitute.ca 
% tkoslowski@perimeterinstitute.ca}
%\thanks{Electronic address: lmodesto@perimeterinstitute.ca}
\affiliation{Perimeter Institute for Theoretical Physics, 31 Caroline St., Waterloo, ON N2L 2Y5, Canada}

\date{\small\today}

\begin{abstract} \noindent
%Essentially there are two aproches to quantum gravity: the perturbative one where the gravitons
%move an a fixed classical background and the opposite non perturbative one where the space-time
%is quantum. 
In this paper we introduce a perturbatively super-renormalizable and unitary theory of quantum gravity
in any dimension $D$. %nonlocal 
In four dimensions the theory is an extension of the Stelle higher derivative 
gravity that involves an infinite number of derivative terms %. quadratic in the curvature tensor. 
%All the extra operators added to the classical Einstein-Hilbert action are 
characterized by two 
entire functions, a.k.a. ``form factors". 
In dimension $D$ we preserve two entire functions and we implement a finite number of local operators
required by the quantum consistency of the theory.
The main reason to introduce the entire functions is to avoid  
%We know that the usual quadratic action is renormalizable 
%but suffers of the unitarity problem because of the presence of a 
ghosts %poltergeist 
(states of negative norm) 
like the one in the four-dimensional Stelle's theory. %in the theory.
The new theory is indeed ghost-free since the two entire functions 
%which does 
have the property to generalize the Einstein-Hilbert action without 
introducing new poles in the propagator. 
By expanding the form factors to the lowest order in a mass scale we introduce,
the local high derivative theory 
is recovered. %expanding the form factors %entire functions 
%to the lowest order in a mass scale 
%introduced in the theory. 
Any truncation of the entire functions gives rise to %produces 
the unitarity violation and it is only by keeping all the infinite series 
that we overcome similar issues. 
%do not fall into these troubles.  
The theory is renormalizable at one loop and finite from two loops upward.
More precisely, the theory turns out to be super-renormalizable because the covariant counter-terms 
have less derivatives then the classical action
and the coefficients of the terms with more
derivatives do not need any kind of infinityÊ
renormalization.Ê
%We analyze the fractal properties of the theory at high energy showing a reduction of the spacetime
%dimension at short scales. 
In this paper we essentially study three classes of form factors, systematically showing the tree-level unitarity. %For the first to the theory is 
%power counting super-renormalizable 

We prove that the gravitation potential is regular in $r = 0$ for all the choices of form factors 
%considered in the paper 
compatible with renormalizability and unitarity. 
We also include Black hole spherical symmetric 
solutions %for two particular choices of the form factors and %are also studied 
omitting higher curvature corrections to the equation of motions. 
For two out of three form factors the solutions are regular and the classical singularity
is replaced by a ``de Sitter-like core" in $r=0$.
%Some preliminary results at 

For one particular %asymptotically polynomial 
choice of the form factors, we prove that the $D$-dimensional ``Newtonian cosmology" is singularity-free
and the Universe %naturally has 
spontaneously follows 
a de Sitter evolution at the ``Planck scale" for any matter content (either dust or radiation). 

%We conclude the article considering 
%Black hole spherical symmetric 
%solutions for some particular choices of the form factors and %are also studied 
%omitting high curvature corrections to the equation of motions. The solutions are regular and the classical singularity
%is replaced by a ``de Sitter-like core" in $r=0$. %Black holes may show a ``multi-horizon" structure 
%depending on the value of the ADM mass.

We conclude the article providing an extensive analysis of the spectral dimension  
for any %dimension 
$D$
and for the three classes of theories. In the ultraviolet regime the spectral dimension takes on different values
for the three cases: 
less than or equal to ``$1$" 
for the first case, ``$0$" for the second one and ``$2$" for the third one.
Once the class of theories compatible with renormalizability and unitarity is defined, 
the spectral dimension has the same short distance ``critical value" or ``accumulation point" for any 
value of the topological dimension $D$. 

\end{abstract}

\maketitle

%\section{Introduction to the theory}

%In this paper we introduc
In 1916 Albert Einstein %introduced an historical 
revolutionized the way physicists thought about gravity with the theory of 
%Here we briefly recall the main ideas underline classical 
``general relativity". This theory works quite well at the classical level, 
but at the theoretical level one of the biggest problems %in theoretical physics 
is to find a theory that is able to reconcile 
general relativity and quantum mechanics. There are  many reasons to believe that 
gravity has to be quantum, some of which are: the quantum nature of matter in the right-hand side of the 
Einstein equations, the singularities appearing in classical solutions of general relativity, and so on. 

Let us recall here the main hypothesis underlying general relarivity.
%The theory is well know to everybody under the name of ``General Relativity".
The grounding principles of ``general relativity" % at the foundation of this theory 
are: % the following: 
(i) gravity is no longer a force as in the Newton's theory, 
but it is the ``curvature of the spacetime", (ii) the symmetry principle underlying the gravitational theory 
is the ``general covariance" or ``invariance under general coordinate transformations",
(iii) the ``energy momentum tensor" for any kind of matter has to be covariantly conserved,
(iv) the dynamics should be described by ``second order" differential equations. 
%with no more then two derivatives.
Assuming these fundamental requirements, we get a ``unique" theory for the dynamics of gravity,
namely the Einstein equations. The theory can be also formulated starting from an action principle
by Einstein and Hilbert.
%
%One of the biggest problems in theoretical physics is to find a theory that is able to reconcile 
%general relativity and quantum mechanics. There are  many reasons to believe that 
%gravity has to be quantum, some of which are: the quantum nature of matter in the right-hand side of the 
%Einstein equations, the singularities appearing in classical solutions of general relativity, etc. 
The action principle we are going to introduce in this paper %is forced to 
makes gravity compatible with
quantum mechanics, as it is the result of 
a synthesis of minimal requirements beyond the classical Einstein-Hilbert action. % we are going to list below.

Let us axiomatically itemize %list 
these requirements one by one: % axiomatically.
\begin{enumerate}
\renewcommand{\theenumi}{(\roman{enumi})}
%\item classical solutions must be singularity-free, 
\item  gravity is still retained as curvature of the spacetime and the underlying symmetry principle 
remains ``general covariance";
\item Einstein-Hilbert action is no longer the correct one, but it should still 
be a good approximation of the theory at an energy scale much smaller than the Planck mass 
or any other invariant scale; % introduced in the theory;
%the action has to be well approximated by the, % Einstein-Hilbert action,
\item solutions of the classical equations of motion must be singularity-free;
\item the theory has to be perturbatively renormalizable or finite at quantum level.
%%%
% The last request is the 
 In other words, we assume ``axiomatically" the validity of the ``perturbative theory".
We claim that any mathematical theory which prides itself to describe the universe 
must be perturbatively consistent. % for one main reason. 
This is empirically true 
for all the other fundamental forces: Weak, Strong and Electromagnetic. % interaction. 
%for atomic physics, nuclear physics, particle physics, quantum mechanics, classical mechanics and general relativity.
If for a general system we find that this is not the case, then we have to change the theory 
or ``dualize" it \cite{d1, d2, d3, d4} 
%before 
instead of trying to solve it at a non-perturbative level; % the theory.
%
%
%%%
\item the spacetime spectral-dimension should decrease with the energy to obtain %in order to have %, at least in principle,
a complete quantum gravity theory in the ultraviolet regime %this hypothesis is 
%\item the theory has to be perturbatively renormalizable at quantum level 
(this hypothesis is strongly related to the previous one (iv)).
%Finiteness and/or renormalizability of the quantized theory. %In oder words we recall 
%the spectral dimension to 
%From now to simplify the notation usually (but not ever) we introduce the following definition 
%$\Box/\Lambda^2 = \Box_{\Lambda}$. 
%Our guiding principle is to find a well defined quantum theory 
%at quantum level 
%with dimensional reduction of the spacetime at high energy.
This property seems to be of ``universal nature" because 
it is common to many approaches to quantum gravity 
\cite{FST1, FST2, FST3, FST4, FST5, FST6, FST7, FST8, FST9, CDT1, CDT2, CDT3}.
The Stelle's theory \cite{Stelle}, the Crane-Smolin theory \cite{CraneSmolin},  
``asimptotically safe quantum gravity" \cite{Reuter, R2, R3, litim, Benedetti, Darione2, Darione3}, ``causal dynamical triangulations" \cite{CDT2, CDT3},``loop quantum gravity" \cite{FST4} and ``string theory" already 
manifest this property with an high energy spectral dimension $d_s = 2$ or less; % \cite{Carlip}. %effective dimension to two.
%However %We consider 
%such reduction is insufficient if we want a unitary theory free from negative norm states.
%We can anticipate that,  
%for the model that we are going to introduce in this paper, the spectral dimension is smaller than one %goes to zero %disappearances 
%in the ultraviolet regime. 
%, % one in this list), 
\item the theory has to be unitary, with no other pole
%poles in the propagator then the graviton 
%degree of freedom %other  than 
%besides 
except 
the graviton one in the propagator;
\item spacetime is a single continuum of space and time and in particular the 
Lorentz invariance is not broken. 
This requirement is supported by recent observations.
%\item The last request is the axiomatic correctness of ``perturbative theory".
%We assume that any mathematical theory that prides itself to describe the universe 
%must be consistent perturbatively. % for one main reason. 
%This is experimentally true 
%for atomic physics, nuclear physics, particle physics, quantum mechanics, classical mechanics and general relativity.
%If we find that this is not the case then we have to change theory or ``dualize" the theory 
%before to try to solve it at a non-perturbative level. % the theory.
 \end{enumerate}
%\item 
%\end{enumerate}
The main hypothesis we abandon with respect to the classical theory %relax 
is the absence of higher derivative terms in the action. As we are going to show we admit indeed 
an infinite tower of operators defined through an ``entire function" of the D'Alembertian differential operator.

%In this paper, firs we generalize the Stelle theory to any dimension and then we restore unitarity. 
%The above hypothesis and t
This work  %its full structure are 
is inspired by papers in four dimensions about a ``nonlocal extension'' of 
gauge theories and gravity published %introduced by Moffat, Cornish and their collaborators 
in the Nineties \cite{efimov, Krasnikov, Moffat1, Moffat2, Moffat3, Moffat4, Moffat5, MoffatG1,MoffatG2}.
%The authors further extended the idea to gravity, having in mind the following logic \cite{MoffatG}. 
%
%The logic is the following: 
For example, 
in \cite{Moffat1, Moffat2, Moffat3, Moffat4, Moffat5, MoffatG1,MoffatG2} %these papers 
the authors considered a modification of the Feynman rules,  
where the coupling constants ($g_i$ for electro-weak interactions and $G_N$ for gravity)
become functions of the momentum $p$. They proved the 
gauge invariance at all orders in gauge theory but only %order by order 
up to the second order in gravity.
For particular choices of $g_i(p)$ or $G_{N}(p)$, the propagators do not show any other pole  
besides those of the standard particle content of the theory, which makes the theory unitary.
On the other hand the theory is also finite if the coupling constants go sufficiently fast to zero 
in the ultra-violet limit. 
%Given this behavior %order by order in the perturbative expansion,
On the basis of these conclusions,  
the problem with gravity remains to find a covariant action that self-contains the properties mentioned 
above: finiteness and/or renormalizability and unitarity.

%phenomenological more facts of general 

%Let us now 
We %are going 
now introduce our theory %, step by step, 
starting from 
%firstly recalling 
the perturbative $D$-dimensional 
``non-renormalizable" Einstein gravity, 
%secondly recalling 
%through high derivatives gravity theories 
and recalling the four dimensional Stelle's \cite{Stelle} %higher-derivative 
quantum gravity,  
which serves as an example 
%will be our first example
of power-counting renormalizable (but not unitary) theory. %,
%to conclude with
%to get to lay the foundations of an
%action which defines a D-dimensional complete quantum gravity theory.

First of all, let us explain why it is important to study quantum gravity in $D$-dimensions.
%The impatient reader can skip to the end of the introduction for the candidate complete 
%quantum gravity bare Lagrangian. 
There are at least five reasons to look for a super-renormalizable theory of gravity 
with extra dimensions. (i) The first reason is that %to work in %one is a technical reason, 
$D>4$ eliminates ``soft-graviton or % we do not have 
infra-red divergences" at quantum level \cite{giddi}; 
(ii) the second one is that only %related to the existence of the ``total cross section" 
in $D\geqslant 7$ there exists a well defined ``total cross section" \cite{giddi};
(iii) the third one is to obtain a well defined ``Kaluza-Klein grand-unification";
(iv) the forth one is the possibility to have a well defined completion  
of the $11$-dimensional supergravity as a candidate for ``M-Theory";
(v) the last reason to study gravity in any spacetime dimension is related to the universality 
of the quantization procedure   
%procedure of quantization 
independently from the number of dimensions ``$D$". 
%Pragmatically 
In synthesis, %we assert: 
we do not want to tune the spacetime dimension to a particular value to make the theory consistent
at quantum level; instead, the theory ``should be" well defined for any value of $D$.

In this paper we use the signature $(+ - \dots -)$. The curvature tensor is defined by 
$R^{\alpha}_{\beta \gamma \delta} = - \partial_{\delta} \Gamma^{\alpha}_{\beta \gamma} + \dots$, 
the Ricci tensor by $R_{\mu \nu} = R^{\alpha}_{\mu \nu \alpha}$, and the curvature scalar by $R = g^{\mu \nu} R_{\mu \nu}$ and $g_{\mu \nu}$, where the metric tensor.

%\section{Introduction}
Perturbative quantum gravity is the quantum theory of a spin-two particle on a fixed 
(conventionally flat) background. %We start 
Starting from the $D$-dimensional Einstein-Hilbert Lagrangian 
\be
{\mathcal L}_{\rm EH} = %- 
\sqrt{ | g |} \, 2 \, \kappa^{-2} \, R \, , 
\ee
 ($\kappa^2 = 32 \pi G_N$) we split 
%introduce a splitting of 
the metric in a background part and in a fluctuation 
\be
%\sqrt{ - g} g^{\mu \nu} =  {g}^{o \mu \nu} + \kappa h^{\mu \nu},
g_{\mu \nu} =  {g}^{o}_{\mu \nu} + \kappa h_{\mu \nu}.
\label{hmunu}
\ee
Using (\ref{hmunu}), 
we then expand the action %that operator 
in powers of the graviton fluctuation $h_{\mu\nu}$ around the fixed background $g^{o}_{\mu \nu}$
that we assume to be the flat Minkowski metric $\eta_{\mu \nu}$. % in the rest of the paper.
Regrettably, the quantum theory is divergent 
at two loops in $D=4$, so we should introduce a counter-term 
proportional to the Riemann tensor at the third power 
\be 
{\mathcal L}_{R^3} \propto G_N \,  \sqrt{ - g} R^{\alpha \beta}_{\gamma \delta} 
R^{\gamma \delta}_{\rho \sigma} 
R^{\rho \sigma}_{\alpha \beta}.
\ee
In the general D dimensional Einstein-Hilbert theory 
the superficial degree of divergence of a Feynman diagram is 
$\delta = L \, D + 2 V -2 I$, where $L$ is the number of loops, % in ther graph,
 $V$ is the number of vertices and $I$ the number of internal lines in the graph.
Substituting in $\delta$ the topological relation %between $V$, $I$ and $L$, 
$L = 1+ I -V$, we obtain
\begin{eqnarray}
\delta = 2 + (D - 2) L \,\,\,\,\, {\rm for} \,\,\,\, {\mathcal L}_{\rm EH} \, . 
\label{Dren}
\ee
In $D>2$ the superficial degree of 
divergence increases with the number of loops. Therefore, we are forced to 
introduce an infinite tower of higher derivative counter-terms and an infinite number of coupling 
constants, thus making the theory not predictive.

%\label{div}
%\end{eqnarray}
Schematically, we can relate the number $L$ of loop divergences %in perturbative quantum gravity 
to 
the counter terms we have to introduce to regularize the theory. In short % in the following way 
%we can write down all the relation between 
\be
S = %-  
\int   {\rm d}^D x \sqrt{|g|} \, \Big[2 \, \kappa^{-2}  \, R + \!\! % \frac{\alpha_{nm}}{\epsilon} 
%\hspace{-0.3cm} 
\underbrace{\,  \sum_{m, n}^{+ \infty}    \frac{\alpha_{nm}}{\epsilon} \,\, \nabla^n R^m}_{n+2 m = 2 +(D -2)L}   
  \Big] ,
  \label{operatorsG}
\ee
where ``$n$" and ``$m$" are integer numbers, $\alpha_{mn}$ are coupling constants and $1/\epsilon$ 
is the cutoff 
in dimensional regularization.

Now we want to comment further % to add %at this point 
%a comment 
about the above statement on the meaning of renormalizability. %before (\ref{operatorsG}). 
Scholars usually claim 
that a theory loses its predictability when %because 
its action consists of an infinity number 
of operators.   Consequently, they believe that the theory can be defined only through  
%we have to do 
an infinite number of measures. % to define the theory.
%Lagrangian. 
Such statement is highly questionable because we should always  
add all the 
possible operators to a Lagrangian
and/or an Hamiltonian describing a physical system. 
%The most important factor we need to worry about 
The most important step we need to take is to assess
%is instead 
whether 
the physical measurable quantities are affected or not  
%
%, but if all the physical measurable quantities are not affected 
by the above-mentioned operators. 
In other words it is essential to establish if such operators are ``relevant o irrelevant" 
to the physical observable. % then 
If we can assume that no physical quantity is susceptible to such operators, we can then empirically infer 
that the coupling constants equal ``zero", as long as 
%iff and only if the theory is unitary. 
%Therefore, 
the theory %must %mandatorily 
satisfies the unitarity requirement. 
%What is really necessary is the unitarity requirement to be satisfied instead of %and not 
%how many operators characterize the theory.
Let us provide an example. Suppose to add other hermitian operators to the standard Hamiltonian of the hydrogen atom.
%standard Hamiltonian. 
We can ``invent" an infinite number of such operators but only few of them 
will be relevant, like for example the relativistic corrections or the Lamb Shift, while all the other ones 
are irrelevant. 
%can be considered 
%are coupled with ``$\sim 0$" 
%coupling constants. 
In this case what is really significant is 
the precision level in the measure of the energy spectrum, compatibly with 
a zero value for the other  couplings, as long as unitarity is satisfied.
%assumption to fix at ``zero" all the other couplings. 
%However, what we really have to take care is unitarity.
%The other main property 
%has to be satisfied is unitarity. % evolution.

A first revolution in quantum gravity in $4D$ was introduced by Stelle \cite{Stelle, Shapirobook} 
with his higher derivative theory 
\be
S = %- 
\int {\rm d}^4 x \sqrt{-g} \Big[ \alpha R_{\mu \nu} R^{\mu \nu} - \beta R^2 + 2 \,  \kappa^{-2} R \Big].
\label{OldStelle}
\ee
If we calculate the upper bound to the superficial degree of freedom for this theory in $D$-dimensions we find 
\be
%&& \hspace{-0.7cm} 
\delta= D L - 4 I + 4 V  
%= D(1+ I - V) - 4 (I-V) \\
%&&  \hspace{-0.5cm}
= D - (D-4)(V-I) \, ,
\label{deltaD}
\ee
so that $\delta = 4$ in $D=4$.
The theory is then 
renormalizable, since all the divergences can be absorbed in the operators already present in 
the Lagrangian (\ref{OldStelle}). Unfortunately however, the propagator contains a physical %poltergeist 
ghost (state of negative norm) that represents  
a violation of unitarity. Probability, as described by the scattering $S$-matrix, is no longer preserved.
Similarly, the classical theory is destabilized, 
since the dynamics can drive the system to become arbitrarily excited,
and the Hamiltonian constraint is unbounded from below. 
On this basis, we can generalize the Stelle theory to a $D$-dimensionl renormalizable one. %theory of gravity generalizing the Stelle theory. 
In short, the Lagrangian with at most $X$ derivatives of the metric is 
\be 
&& \hspace{-0.5cm} {\mathcal L}_{D-{\rm Ren} } = a_1 R + a_2 R^2 + b_2 R_{\mu \nu}^2 + %\dots 
\label{localDren}\\
&& \hspace{-0.5cm} \dots + a_x R^{X/2} + b_x R_{\mu \nu}^{X/2} +
c_X R_{\mu \nu \rho \sigma}^{X/2} + d_X R \, \Box^{\frac{X}{2} - 2} R \dots \, .\nonumber 
\ee
In the second line,
the dots on the left imply a finite number of extra 
terms with fewer derivatives of the metric tensor, and the dots on the right indicate 
a finite number of operators with the same 
number of derivatives but higher powers of the curvature ($O(R^2 \Box^{X/2-4} R)$).

In this theory, the power counting tells us that the maximal superficial degree of divergence 
of a Feynmann graph is 
\be
\delta = D - (D - X)(V-I).
\label{DDX}
\ee
%and f
For $X = D$ the theory is renormalizable since the maximal divergence is $\delta =D$
and all the infinities can be absorbed in the operators already present in the action (\ref{localDren}).

%We can define t
The general action of ``derivative order $N$" can be found combining curvature tensors and covariant derivatives of the curvature tensor. In short the action reads %in terms of the number of derivatives of the metric tensor 
as follows \cite{shapiro},
\be
\!\!\!\! S = \sum_{n=0}^{N+2} \alpha_{2 n} \Lambda^{D - 2 n} \int d^D x \sqrt{ | g |} \, \mathcal{O}_{2 n}  (\partial_{\rho } g_{\mu \nu}) + S_{\rm NL} \, ,
\label{generalAction}
\ee
where $\Lambda$ is a mass scale in our fundamental theory, $\mathcal{O}_{2n} (\partial_{\rho} g_{\mu \nu})$ denotes the general covariant scalar term containing 
``$2 n$" derivatives of the metric $g_{\mu \nu}$, while $S_{\rm NL}$ is a nonlocal action term that 
we are going to set %define 
later \cite{NL1, NL2, NL3, NL4}. The maximal number of derivatives in the local part of the action is $2 N +4$.
We can classify the local terms in the following way,
\be
&& N= 0 \,\, : \,\,\, S_{0} = \lambda + c_0^{(0)} R + c_1^{(0)} R^2 + c_2^{(0)} R_{\mu \nu} R^{\mu \nu} \, , \nonumber \\
&& N = 1 \,\, : \,\,\,S_{1} = S_{0} + 
 c_1^{(1)} R^3_{\dots}  + c_2^{(1)} \nabla R_{\dots} \nabla R_{\dots} \, , \nonumber \\
&&N= 2 \,\, : \,\,\,
 S_{2} = S_{0} + S_{1} + 
 c^{(2)}_1 R^4_{\dots} \nonumber \\
 && \hspace{1.5cm} 
+ c^{(2)}_2 R_{\dots} \nabla R_{\dots} \nabla R_{\dots} + c_3^{(3)} \nabla^2 R_{\dots} \nabla^2 R_{\dots} \, , 
\nonumber \\
&& \dots \nonumber \\ %\dots \dots \dots \, , \\
&& \dots \nonumber \\
%&& \dots \\
&& N = N \,\, : \,\,\,S_{N} = \!\!\!\! \!\!\!
\sum_{i=0}^{N-1, \, N>0} \!\!\!\!\!\!  S_{i} \, + \,  c_1^{(N)} R^{N+2}_{\dots} + \nonumber \\
&& + c_2^{(N)} R^{N-1}_{\dots} \nabla R_{\dots} \nabla R_{\dots} + \dots + c^{(N)}_M R_{\dots} \Box^N R_{\dots} \,\,  .
\ee
In the local theory (\ref{localDren}), renormalizability %of the local theory 
requires $X = D$, so that the relation between
the spacetime dimension and the derivative order is $2N +4 = D$.
To avoid fractional powers of the D'Alembertian operator, we take
$2 N + 4 = D_{\rm odd} +1$ in odd dimensions and $2 N + 4 = D_{\rm even}$
in even dimensions.

In this paper, we are focused on the renormalizability and unitarity of the theory,  
so the main quantity to calculate is the graviton propagator.
Although 
the action is complicated, we only need to consider 
the quadratic operators in the curvature to get the ``two points function".
Given the general structure (\ref{generalAction}), for $N>0$ and $n>2$ 
contributions to the propagator come only from the following operators, %terms of the second order in the 
%curvature i.e. 
\be
%c_{1}^{(n)} 
R_{\mu \nu} \Box^{n-2} R^{\mu \nu} \, , \,\,\, %+ c_{2}^{(n)} 
R  \Box^{n-2} R \, , \,\,\, %+ c_3^{(n)} 
R_{\mu \nu \alpha \beta } \Box^{n-2} R^{\mu \nu \alpha \beta}.
\label{terms}
\ee
However, using the Bianchi and Ricci identities one can reduce the terms listed above from three 
to two (with total $2n$ derivatives)
\cite{sg},  
\be
&& \hspace{-0.5cm} R_{\mu \nu \alpha \beta } \Box^{n-2} R^{\mu \nu \alpha \beta}    
\label{property} \\
&& \hspace{-0.5cm} = - \nabla_{\lambda} R_{\mu \nu \alpha \beta } \Box^{n-3} \nabla^{\lambda} R^{\mu \nu \alpha \beta} 
+  O(R^3) + \nabla_{\mu} \Omega^{\mu} \nonumber \\
&& \hspace{-0.5cm} = 4  R_{\mu \nu} \Box^{n-2} R^{\mu \nu} 
- R  \Box^{n-2} R + O(R^3) + \nabla_{\mu} \Omega^{\prime \, \mu} \, , \nonumber 
\ee
where %$a, b$ are numerical coefficients 
$\nabla_{\mu} \Omega^{ \mu}$ 
and $\nabla_{\mu} \Omega^{\prime \, \mu}$ are total divergence terms.
Applying 
(\ref{property})
to (\ref{terms}), for $n>2$ we discard the third term and we keep the first two. 

We now have to define the ``non-local" action term in (\ref{generalAction}). 
As we are going to show, both super-renormalizability and unitarity require the following 
two non-local operators,
\be 
 R_{\mu \nu} \,  h_2( - \Box_{\Lambda}) R^{\mu \nu} \, , \,\,\, \, 
  R   \, h_0( - \Box_{\Lambda}) \, R \, .
\ee 
The full action, focusing mainly on the non-local terms and the quadratic part in the curvature, reads
\be 
&& %\hspace{0.0cm} 
S = \int d^D x \sqrt{|g|} \Big[2\, \kappa^{-2} \, R + \bar{\lambda}
\nonumber \\
&& + \sum_{n=0}^{N} \Big( 
a_n \, R \, (-\Box_{\Lambda})^n \, R  + 
b_n \, R_{\mu \nu} \, (-\Box_{\Lambda})^n \, R^{\mu \nu} 
%a_n \, R \, (-\Box_{\Lambda})^n \, R 
\Big) 
\nonumber \\
&& %\hspace{0cm} 
%+ R_{\mu \nu} \, h_2( - \Box_{\Lambda}) \, R^{\mu \nu} +
+ R  \, h_0( - \Box_{\Lambda}) \, R +
R_{\mu \nu} \, h_2( - \Box_{\Lambda}) \, R^{\mu \nu}
 \, \nonumber \\
&& %\hspace{0cm}
+ \underbrace{\dots\dots\dots O(R^3) \dots\dots\dots + R^{N+2}}_{\mbox{Finite number of terms}} \Big] 
 \, . \label{action}
\ee 
The last line is a collection of local terms that are renormalized at quantum level.
In the action, 
 the couplings and the non-local functions have the following dimensions: $[a_n] =[b_n]= M^{D-4}$, $[\kappa^2] = M^{2-D}$,
 $[\bar{\lambda}]=M^D$, $[h_2] = [h_0] = M^{D-2}$.
 
At this point, we are ready to expand the Lagrangian at the second order in the graviton fluctuation.
Splitting the spacetime metric in the flat Minkowski background and the fluctuation $h_{\mu \nu}$
as defined in (\ref{hmunu}), we get \cite{HigherDG}
\be
&& \hspace{-0.3cm} \mathcal{L}_{\rm lin} = %\frac{1}{2} h^{\mu \nu} \mathcal{O}_{\mu \nu, \rho \sigma} h^{\rho \sigma}
- \frac{1}{2} [ h^{\mu \nu} \Box h_{\mu \nu} + A_{\nu}^2 + (A_{\nu} - \phi_{, \nu})^2 ] \nonumber \\
&&  \hspace{-0.4cm} + \frac{1}{4} \Big[ \frac{\kappa^2}{2} \Box h_{\mu \nu}  \beta( \Box) \Box h^{\mu \nu} 
- \frac{\kappa^2}{2} A^{\mu}_{, \mu}  \beta( \Box) A^{\nu}_{, \nu} 
 \nonumber \\
&&  \hspace{-0.4cm} - \frac{\kappa^2}{2} F^{\mu \nu}  \beta( \Box) F_{\mu \nu} 
+ \frac{\kappa^2}{2} (A^{\alpha}_{, \alpha} - \Box \phi) \beta( \Box) (A^{\beta}_{, \beta} - \Box \phi)
\nonumber \\
&& \hspace{-0.4cm} + 2 \kappa^2 \left(A^{\alpha}_{, \alpha}  - \Box \phi \right) \alpha( \Box) (A^{\beta}_{, \beta} - \Box \phi ) \Big] \,  ,
\label{quadratic} 
\ee
where $A^{\mu} = h^{\mu \nu}_{\,\,\,\, , \nu}$, $\phi = h$ (the trace of $h_{\mu \nu}$), 
$F_{\mu \nu} = A_{\mu , \nu} - A_{\nu, \mu}$ and the functionals of the D'Alembertian operator 
$\beta (\Box), \alpha(\Box)$ are defined by 
\be
&&\alpha(\Box)/2 :=  \sum_{n = 0}^{N} a_n ( - \Box_{\Lambda})^n + h_0(- \Box_{\Lambda}) \, , \nonumber \\
&& \beta(\Box)/2 :=  \sum_{n = 0}^{N} b_n ( - \Box_{\Lambda})^n + h_2(- \Box_{\Lambda}) \,  .
\label{alphabeta}
\ee
The d'Alembertian operator in (\ref{quadratic}) and (\ref{alphabeta}) must be conceived on %relative 
%to 
the flat spacetime. %Minkowski background. 
The linearized Lagrangian (\ref{quadratic}) is invariant under infinitesimal coordinate transformations 
$x^{\mu} \rightarrow x^{\mu} + \kappa \xi^{\mu}(x)$, where $\xi^{\mu}(x)$ is an infinitesimal vector field 
of dimensions $[\xi(x)] = M^{(D-4)/2}$. Under this transformation, the graviton field turns into 
\be 
h_{\mu \nu} \rightarrow h_{\mu \nu} - \xi(x)_{\mu,\nu} - \xi(x)_{\nu,\mu}.
\label{gaugetra} 
\ee
The presence of the local gauge simmetry (\ref{gaugetra}) 
calls for the addition of a gauge-fixing term
to the linearized 
Lagrangian (\ref{quadratic}). Hence, we choose the following fairly general gauge-fixing operator
\be
&& \mathcal{L}_{\rm GF} = \lambda_1 (A_{\nu} - \lambda \phi_{,\nu}) \omega_1(-\Box_{\Lambda}) (A^{\nu} - \lambda \phi^{,\nu})
\nonumber \\
&& + \frac{\lambda_2 \, \kappa^2}{8} (A^{\mu}_{,\mu} - \lambda \Box \phi) \beta(\Box) \omega_2(-\Box_{\Lambda}) (A^{\nu}_{,\nu} - \lambda \Box \phi) \nonumber \\
&& + \frac{\lambda_3 \, \kappa^2}{8}    
F_{\mu \nu} \,
\beta(\Box) \omega_2(-\Box_{\Lambda}) \, 
 F^{\mu \nu}  \, ,
 \label{GF}
\ee
where $\omega_i( - \Box_{\Lambda})$ are three weight functionals \cite{Stelle, NL4}.
In the harmonic gauge %is common to define $\lambda_1 = 1/\zeta$
$\lambda=\lambda_2 = \lambda_3 = 0$ and %it 
%is common to define 
$\lambda_1 = 1/\xi$.
The linearized gauge-fixed Lagrangian reads 
\be
\mathcal{L}_{\rm lin} + \mathcal{L}_{\rm GF} = 
\frac{1}{2} h^{\mu \nu} \mathcal{O}_{\mu \nu, \rho \sigma} h^{\rho \sigma},
\label{O}
\ee
where the operator %inverse of the propagator 
$\mathcal{O}$ is made of two terms, one coming from the linearized 
Lagrangian 
(\ref{quadratic}) and the other from the gauge-fixing term(\ref{GF}).
Inverting the operator $\mathcal{O}$ \cite{HigherDG} we find the %following 
two point function in the harmonic gauge ($\partial^{\mu} h_{\mu \nu} = 0$),
\be
&& \hspace{-1.2cm}  \mathcal{O}^{-1}(k) = \frac{\xi (2P^{(1)} + \bar{P}^{(0)} ) }{2 k^2 \, \omega_1( k^2/\Lambda^2)} 
%&& \hspace{-0.4cm} 
+ \frac{P^{(2)}}{k^2 \Big(1 + \frac{k^2 \kappa^2 \beta(k^2)}{4} \Big)} 
\nonumber \\
&&
- \frac{P^{(0)}}{2 k^2 \Big( \frac{D-2}{2} - \frac{D \beta(k^2) \kappa^2/4 + (D-1) \alpha(k^2) \kappa^2}{2} \Big) } \,  . \label{propagator}
 %\\
%&& \hspace{-0.4cm} 
%+ \frac{\zeta}{2 k^2 \, \omega_1( k^2/\Lambda^2) (2P^{(1)} + \bar{P}^{(0)}. }
%\nonumber \\
\ee
The tensorial 
indexes for the operator $\mathcal{O}^{-1}$ and the projectors $P^{(0)},P^{(2)},P^{(1)},\bar{P}^{(0)}$ have been omitted and the functions $\alpha(k^2)$ and $\beta(k^2)$ are achieved by replacing $-\Box \rightarrow k^2$ in the definitions (\ref{alphabeta}). The projectors are defined by %We have also introduced the following projectors 
\cite{HigherDG, VN}
\be
 && P^{(2)}_{\mu \nu \rho \sigma}(k) = \frac{1}{2} ( \theta_{\mu \rho} \theta_{\nu \sigma} +
 \theta_{\mu \sigma} \theta_{\nu \rho} ) - \frac{1}{D-1} \theta_{\mu \nu} \theta_{\rho \sigma}  ,
 \nonumber \\
 && P^{(1)}_{\mu \nu \rho \sigma}(k) = \frac{1}{2} \left( \theta_{\mu \rho} \omega_{\nu \sigma} +
 \theta_{\mu \sigma} \omega_{\nu \rho}  + 
 \theta_{\nu \rho} \omega_{\mu \sigma}  +
  \theta_{\nu \sigma} \omega_{\mu \rho}  \right) ,
 \nonumber \\
% \ee
 %\be
 && P^{(0)} _{\mu\nu\rho\sigma} (k) = \frac{1}{D-1}  \theta_{\mu \nu} \theta_{\rho \sigma} \,\, , \hspace{0.1cm}
\bar{P}^{(0)} _{\mu\nu\rho\sigma} (k) =  \omega_{\mu \nu} \omega_{\rho \sigma}, \nonumber \\
&& \bar{\!\bar{P}}^{(0 )} _{\mu\nu\rho\sigma}  = \theta_{\mu \nu} \omega_{\rho \sigma}
+ \omega_{\mu \nu} \theta_{\rho \sigma} \, , \nonumber \\
&& \theta_{\mu \nu} = \eta_{\mu \nu} - \frac{k_{\mu} k_{\nu}}{k^2} \,\, , \,\,\,
 \omega_{\mu \nu} = \frac{k_{\mu} k_{\nu}}{k^2} \, .
\label{proje}
\ee
%where $\theta_{\mu \nu} = \eta_{\mu \nu} - k_{\mu} k_{\nu}/k^2 \,\, , \, 
 %\omega_{\mu \nu} = k_{\mu} k_{\nu}/k^2$.
%\section{Power counting renormalizability}
%\section{ %Entire functions and 
% Renormalizability}
%

%In this section 
%At this point we want 
%The goal now is to find an upper bound to the divergences for our multidimensional theory of  
%gravity; before doing 
%this, we have to construct the entire functions $h_2$ and $h_0$.
Looking at the last two gauge invariant terms in (\ref{propagator}), we deem convenient to introduce the following definitions, 
\be
&& \hspace{-0.4cm}\bar{h}_2(z) = 1 + \frac{\kappa^2 \Lambda^2}{2}  z \sum_{n=0}^N b_n z^n + \frac{\kappa^2 \Lambda^2}{2}
z \, h_2(z) \, , \label{barh2h0}\\
&& \hspace{-0.4cm}\left(\frac{D-2}{2} \right)\bar{h}_0(z) =  \frac{D-2}{2} - \frac{\kappa^2 \Lambda^2 D}{4}  z 
\Big[\sum_{n=0}^N b_n z^n + h_2(z) \Big] \nonumber \\
&& \hspace{2.8cm}- \kappa^2 \Lambda^2 (D - 1)  z \Big[\sum_{n=0}^N a_n z^n + h_0(z) \Big] , \nonumber 
\ee
where again $z = - \Box_{\Lambda}$. 
Through the above definitions (\ref{barh2h0}),
the gauge invariant part of the propagator greatly simplifies to
\be
 \mathcal{O}^{-1}(k)^{\xi = 0} = \frac{1}{k^2}
\left( \frac{P^{(2)}}{\bar{h}_2} 
- \frac{P^{(0)}}{(D-2) \bar{h}_0 } \right).
\label{propgauge}
\ee
 
% and after 
%we had already done all the above considerations.
%The author of the paper is Tomboulis and we will follow all th
%The theory developed in \cite{Tombo}  i
We clarify now the incompatibility of the unitarity with a polynomial choice of either $\bar{h}_0, \bar{h}_2$ or,
equally, % equivalently 
%a polynomial choice 
of the two functions 
$\alpha(\Box), \beta(\Box)$.
If we assume for a moment $\alpha(\Box)$ and $\beta(\Box)$ to be polynomial, then each of the two %gauge invariant 
fractions in the propagator takes
%in short, %the following polynomial  = p_n(x)$, where $p_n(x)$ is a polynomial of degree $n$, 
%In this case, as it will be evident in the next section, 
%the propagator takes 
the following form, 
%\vspace{-0.3cm}
\be
\frac{1}{k^2(1+ p_n(k^2))} = \frac{c_0}{k^2} + \sum_i \frac{c_i}{k^2 - M_i^2},
\label{poli}
%\vspace{0.2cm}
\ee
\vspace{0.1cm}\\
where $p_n(x)$ is a polynomial of degree $n$.
In (\ref{poli}) we used the factorization theorem for polynomials and the partial fraction 
decomposition \cite{NL4}. When multiplying the left and right side of (\ref{poli}) by $k^2$
and 
considering 
the ultraviolet limit $k^2 \rightarrow + \infty$, we find that at least one of the coefficients $c_i$ is negative. Therefore 
the theory contains at least a ghost 
%poltergeist 
in the spectrum. 
The conclusion is that $h_2, h_0$ cannot be polynomial.

Once established that $h_2$ and $h_0$ are not polynomial functions, we demand 
%proceed with the following demands on $\bar{h}_2$ and $\bar{h}_0$. 
%Considering \cite{Tombo}, 
%According to the structure of the functions $\bar{h}_i(z)$ ($i=0,2$)) in (\ref{barh2h0}), we require 
the following general properties for the transcendental entire functions $h_i(z)$ ($i = 0,2$) and/or 
$\bar{h}_i(z)$ ($i = 0,2$)
\cite{NL4}:
\begin{enumerate}
\renewcommand{\theenumi}{(\roman{enumi})}
\renewcommand{\labelenumi}{\theenumi}
\item $\bar{h}_i(z)$ ($i=0,2$) is real and positive on the real axis and it has no zeroes on the 
whole complex plane $|z| < + \infty$. This requirement implies that there are no 
gauge-invariant poles other than the transverse massless physical graviton pole.
\item $|h_i(z)|$ has the same asymptotic behavior along the real axis at $\pm \infty$.
\item There exists $\Theta>0$ such that 
\be
&& \lim_{|z|\rightarrow + \infty} |h_i(z)| \rightarrow | z |^{\gamma + N}, \nonumber \\
%\,\, , \,\,\,\,  
&& \gamma\geqslant D/2 \,\,\,\, {\rm for} \,\,\,\, D = D_{\rm even} \, , \nonumber \\
&& \gamma\geqslant (D-1)/2 \,\,\,\, {\rm for} \,\,\,\, D = D_{\rm odd} \, ,
\label{tombocond}
\ee 
for the argument of $z$ in the following conical regions  
\be
&& \hspace{-0.2cm} 
C = \{ z \, | \,\, - \Theta < {\rm arg} z < + \Theta \, , \,\,  \pi - \Theta < {\rm arg} z < \pi + \Theta \} , \nonumber \\
&&  \hspace{-0.2cm} 
{\rm for } \,\,\, 0< \Theta < \pi/2. \nonumber 
\ee
This condition is necessary in order to achieve the (supe-)renormalizability of the theory that we 
are going to show here below. The necessary 
asymptotic behavior is imposed not only on the real axis, (ii) but also in the conic regions that surround it.  
In an Euclidean spacetime, the condition (ii) is not strictly necessary if (iii) applies.
\end{enumerate}
%Given the above properties, 
Let us then examine the ultraviolet behavior of the quantum theory.
According the property (iii) in the high energy regime, the propagator in the momentum space goes as 
$$\mathcal{O}^{-1}(k) \sim 1/k^{2 \gamma +2N +4} \,\,\,\, {\rm for \,\, large} \,\,\,\, k^2 $$ 
(see (\ref{action}, \ref{barh2h0}, \ref{propgauge})).
However, the $n$-graviton interaction has the 
same leading scaling of the kinetic term, since it can be written in the following schematic way,
\be
&& \hspace{-0.25cm}
{\mathcal L}^{(n)} \sim  h^n \, \Box_{\eta} h \,\,  h_i( - \Box_{\Lambda}) \,\, \Box_{\eta} h \nonumber \\
&& \hspace{0.5cm}
\rightarrow h^n \, \Box_{\eta} h 
\,  ( \Box_{\eta} + h^m \, \partial h \partial )^{\gamma + N} \, 
\Box_{\eta} h , 
\label{intera}
\ee
where the indexes for the graviton fluctuation $h_{\mu \nu}$ are omitted 
and $h_i( - \Box_{\Lambda})$ is the entire function defined by the properties (i)-(iii). 
From (\ref{intera}), the superficial degree of divergence in a spacetime of ``even" dimension is 
\be
%&& 
&&\hspace{-0.8cm} 
\delta_{\rm even} = D_{\rm even} L - (2 \gamma + 2N+ 4) I + (2 \gamma + 2N+ 4) V \nonumber \\
&& = D_{\rm even} L - (2 \gamma + D_{\rm even}) I + (2 \gamma + D_{\rm even}) V \nonumber \\ %&& \hspace{0.42cm} 
&& = D_{\rm even} - 2 \gamma  (L - 1).
\label{diver}
\ee
On the other hand, in a spacetime of ``odd" dimension the upper limit to the degree of divergence is
%&&\hspace{-0.8cm} 
\be
\delta_{\rm odd} = D_{\rm odd} - (2 \gamma+1)  (L - 1).
\label{diverodd}
\ee
In (\ref{diver}) and (\ref{diverodd}) we used again the topological relation between vertexes $V$, internal lines $I$ and 
number of loops $L$: $I = V + L -1$. %For the choice we have already  introduced in the previous 
%section 
%We see that only 1-loop diagram are divergent if $\gamma > 2$ and the theory is super-renormalizable.
Thus, if $\gamma > D_{\rm even}/2$ or $\gamma > (D_{\rm odd}-1)/2$, only 1-loop divergences exist and the theory is super-renormalizable\footnote{
A ``local" super-renormalizable quantum gravity with a large 
number of metric derivatives was for the first time introduced in \cite{shapiro}.
}.
Only a finite number of constants are renormalized in the action (\ref{action}), i.e. 
$\kappa$, $\bar{\lambda}$, $a_n$, $b_n$ and the finite number of couplings that multiply the operators in 
the last line. % of the action. %(\ref{action}) 
 %the quantities 
%$\beta$, $\beta_2$, $\beta_0$ and eventually the cosmological constant are renormalized, 
The renormalized action reads
\be 
&& %\hspace{0.0cm} 
 \hspace{-0.4cm} S= \int d^D x \sqrt{|g|} \Big[2\, Z_{\kappa} \, \kappa^{-2} \, R + Z_{\bar{\lambda}} \bar{\lambda}
\nonumber \\
&& \hspace{-0.4cm} + \sum_{n=0}^{N} \Big( 
Z_{a_n} \, a_n \, R \, (-\Box_{\Lambda})^n \, R  + 
Z_{b_n} \, b_n \, R_{\mu \nu} \, (-\Box_{\Lambda})^n \, R^{\mu \nu} 
%a_n \, R \, (-\Box_{\Lambda})^n \, R 
\Big) 
\nonumber \\
&& \hspace{-0.4cm} 
%+ R_{\mu \nu} \, h_2( - \Box_{\Lambda}) \, R^{\mu \nu} +
+ R  \, h_0( - \Box_{\Lambda}) \, R +
R_{\mu \nu} \, h_2( - \Box_{\Lambda}) \, R^{\mu \nu}
 \, \nonumber \\
&& \hspace{-0.4cm}
+ %\underbrace{
Z_{c_1^{(1)}} c_1^{(1)} \, R^3 +  \dots\dots\dots + Z_{c_1^{(N)}} \, c_1^{(N)} \, R^{N+2}
%}_{\mbox{Finite number of terms}} \Big] 
 \, \Big]. \label{actionRen}
\ee 
All the couplings in (\ref{actionRen}) must be understood as renormalized at an energy scale $\mu$. 
Contrarily,   
the functions $h_i(z)$ are not renormalized. To understand this point thoroughly, we can write 
the generic entire functions as a series, i.e. $h_i(z) = \sum_{r=0}^{+\infty} a_r z^r$.
Because of the superficial degrees of divergence (\ref{diver}) and (\ref{diverodd}), 
there are no counterterms that renormalize $a_r$ for $r > N$. As a matter of fact, 
the couplings in the second line of (\ref{actionRen}) already incorporate the renormalizations of 
the coefficients $a_r$ for $r \leqslant N$.
%already 
 %Only the 
%coefficients $a_r$ for $r \leqslant N$ could be renormalized; however, 
%the other couplings in (\ref{actionRen}) already incorporate 
%such renormalization.
 Therefore,
the non-trivial dependence of the entire functions $h_i(z)$ on their argument is preserved at quantum level.

Imposing the conditions (i)-(iii) we have the freedom to choose the following form for the functions 
$h_i$,
\be
&& \hspace{-0.5cm} h_2(z) = \frac{ V(z)^{-1} -1 - \frac{\kappa^2 \Lambda^2}{2} \, z \sum_{n=0}^N \tilde{b}_n \, z^n}{\frac{\kappa^2 \Lambda^2}{2}\, z} \, , \nonumber \\
&& \hspace{-0.5cm} h_0(z) = - \frac{V(z)^{-1} -1 + \kappa^2 \Lambda^2 \, z \sum_{n=0}^N \tilde{a}_n \, z^n}{
\kappa^2 \Lambda^2 \, z} \, , 
\label{hz}
\ee
for general parameters $\tilde{a}_n$ and $\tilde{b}_n$. 
%In general in order to be compatible with the conditions (i)-(iii), we have to take $\bar{h}_i(z) = \alpha \, \exp H_i(z)$, where
%$H_i(z)$ ($i=2,0$)  
Here $V(z)^{-1}= e^{H(z)}$ and $H(z)$ 
is an entire function that exhibits logarithmic asymptotic behavior in the conical region $C$. 
%Since $H(z)$ is an entire function, 
%$\bar{h}_i(z)$ 
%$\exp H(z)$ 
The form factor 
$\exp H(z)$ has no zeros in the entire complex plane for $|z|< + \infty$. % as required by the property (iii).
Furthermore, the non-locality in the action is actually a ``soft" form of non locality, because %$\exp H(z)$ %\bar{h}_i(z)$ 
%is an exponential 
%function and 
a Taylor expansion of $h_i(z)$ eliminates the denominator $\Box_{\Lambda}$ at any energy scale.
%For this reason, the entire functions $h_i(z)$ can be called {\em quasi polynomial}.

%Again following \cite{Tombo} t
%In the rest of the paper we assume $H_2(z) = H_0(z) \equiv H(z)$. 
The entire function $H(z)$, which is compatible with the property (iii), 
can be defined as %in the following way 
\be
H(z) = \int_0^{p_{\gamma + N + 1 }(z)} \frac{1 - \zeta(\omega)}{\omega} {\rm d} \omega \, , 
\label{Hz}
\ee
where $p_{\gamma +N+1}(z)$ and $\zeta(z)$ must 
satisfy the following requirements:
\begin{enumerate}
\renewcommand{\theenumi}{\alph{enumi}.}
\renewcommand{\labelenumi}{\theenumi}
\item $p_{\gamma +N+1 }(z)$ is a real polynomial of degree $\gamma+N+1$ 
with $p_{\gamma +N+1}(0) = 0$,
\item $\zeta(z)$ is an entire and real function on the real axis with $\zeta(0) = 1$,
\item $|\zeta(z)| \rightarrow 0$ for $|z| \rightarrow \infty$ in the conical region $C$ defined in (iii). 
\end{enumerate}

Let us investigate the unitarity of the theory.
We assume that the theory is renormalized at some scale $\mu_0$. 
%If we want that 
If we set  
\be
\tilde{a}_n = a_n(\mu_0) \,\, , \,\,\,\, \tilde{b}_n = b_n(\mu_0),
\label{betaalpha}
\ee
the bare propagator does not possess other gauge-invariant pole than %other then 
the physical graviton one.
Since the energy scale $\mu_0$ is taken as the renormalization point, we get  %we find 
%together with 
$\bar{h}_2 = \bar{h}_0 = V(z)^{-1} = \exp H(z)$. 
Thus, only the physical massless spin-2 graviton pole
occurs in the bare propagator and (\ref{propgauge}) reads 
\be 
 \mathcal{O}^{-1}(k)^{\xi = 0} = \frac{V(k^2/\Lambda^2)  } {k^2}
\left( P^{(2)} 
- \frac{P^{(0)}}{D-2 }  \right).
\label{propgauge2}
\ee
%Where asymptotycallly $\bar{h}(z) \rightarrow z^{\gamma+1}$ and in the momentum space 
% $\bar{h}(k^2) \rightarrow  k^{2 \gamma +2}$. 
%Suppose 

The momentum or energy scale at which the relation between the quantity computed and the quantity 
measured is identified is called the subtraction point and is indicated usually by ``$\mu$" \cite{Hatfield}. 
The subtraction point is arbitrary and unphysical, so the final answers do not have to depend 
on the subtraction scale $\mu$. Therefore, the derivative $d/d \mu^2$ of physical quantities has to be zero.
In our case, if we choose another renormalization scale $\mu$, then the bare propagator acquires poles. 
However, these poles 
cancel in the dressed physical propagator because the shift in the bare part is cancelled 
by a corresponding shift in the self energy. 
%%%
The renormalized action (\ref{actionRen}) will produce finite Green's functions to whatever order 
in the coupling constants we have renormalized the theory to. For example, the $2$-point Green's 
function at the first order 
in the couplings $a_n$, $b_n$ %(non considering the tensorial structure 
can be schematically written as
\be
%\mathcal{O}^{-1} = \frac{e^{-H(k^2/\Lambda^2)}}{k^2 - \Sigma_R(k^2)}
G^{-1}_{2 R} = V(k^2/\Lambda^2) \, (k^2 - \Sigma_R(k^2)),
\ee
where the renormalization prescription requires that $\Sigma_R$ satisfies (on shell)
\be
\Sigma_R(0) = 0 \,\,\,\,\,\, {\rm and} \,\,\,\,\,\, \frac{\partial \Sigma_R}{\partial k^2} \Big|_{k^2 = 0} =0.
\ee
%%%
We did not consider the tensorial structure and the longitudinal components because 
they project away %by gauge invariance
when attached to a conserved energy tensor. 

The subtraction point is arbitrary and therefore we can take the renormalization prescription 
off shell to $k^2 = \mu^2$. The couplings we wish to renormalize must be dependent 
on the chosen subtraction point,
$a_n(\mu)$ and $b_n(\mu)$, in such a way that the experimentally 
measured couplings do not vary on shell.   
The renormalized Green's function $G^{-1}_{2 R}$ at $\mu^2$
must produce the same 
Green function when expressed in terms of bare quantities.
Consequently, 
the scalings $Z_{a_n}$ and $Z_{b_n}$ must also depend on  
$\mu^2$. 
The Green's function written in terms 
of bare quantities can not depend on $\mu^2$, but since $\mu^2$ is arbitrary, the renormalized Green's 
function must not either. This fact,
\be
\frac{d G^{-1}_{2 R} }{d \mu^2} = 0 \, ,
\ee 
% follows from 
is well known as the renormalization group invariance.
 
When $h_2(z)=h_0(z)=0$, the action in (\ref{action}) reduces to (\ref{localDren}) and fails to be unitary.
Unitarity can be achieved only if 
$$a_n= b_n = \dots = d_n = \dots =0 \,\,\,\, {\rm in} \,\,\,\, (\ref{localDren}) \,  , 
$$ which corresponds 
to the D-dimensional Einstein-Hilbert non renormalizable action.

%The same procedure is not applicable to the case $h_2(z)=h_0(z)=0$, because 
%the theory fails to be renormalizable when the unitarity requirement %$\beta_2=\beta_0=0$ 
%is imposed 
%and an infinite tower of counterterms has to be added to the action.

An explicit example of $\exp H(z)$ that satisfies the properties (i)-(iii) can be easily constructed. 
There are of course many ways to choose $\zeta(z)$, but we focus here on the exponential choice  
$\zeta(z) = \exp(- z^2)$, which satisfies requirement c. outlined for (\ref{Hz}) in a conical region $C$ with 
$\Theta =\pi/4$. 
The entire function $H(z)$ is the result of the integral defined in (\ref{Hz})
\be
&& \hspace{-0.3cm} 
H(z) = \frac{1}{2} \left[ \gamma_E + 
\Gamma \left(0, p_{\gamma+N+1}^{2}(z) \right) \right] + \log \left( p_{\gamma+N+1}(z) \right) \, , 
\nonumber \\
&& \hspace{-0.3cm} {\rm Re}( p_{\gamma+N+1}^{2}(z) ) > 0, 
\label{HD}
\ee
where $\gamma_E=0.577216$ is the Euler's constant and  
$\Gamma(a,z) = \int_z^{+ \infty} t^{a -1} e^{-t} d t$ is the incomplete gamma function.  
If we choose $p_{\gamma+N+1}(z) = z^{\gamma +N+ 1}$, $H(z)$ simplifies to:
\be
&& %\hspace{-0.5cm} 
H(z) = \frac{1}{2} \left[ \gamma_E + \Gamma \left(0, z^{2 \gamma +2 N+2} \right) \right] + \log (z^{\gamma +N+1}) \, ,
\nonumber \\
&& {\rm Re}(z^{2 \gamma +2N+2}) > 0. 
\label{H0}
\ee
Another equivalent expression for the entire function $H(z)$ is given by the following series
\be
&& H(z) = \sum_{n =1}^{+ \infty} ( -1 )^{n-1} \, \frac{p_{\gamma +N+1}(z)^{2 n}}{2n \, n!} , \nonumber \\
&& {\rm Re}( p_{\gamma+N+1}^{2}(z) ) > 0.
\label{HS}
\ee
%%%%%
%%%%%
For $p_{\gamma +N+1}(z) = z^{\gamma +N+1}$ the $\Theta$ angle, which defines the cone $C$,  
is $\Theta = \pi/(4 \gamma +4 N + 4)$. 
According to the above expression (\ref{HS}) we find the following behavior near $z = 0$ for the 
particular choice $p_{\gamma+N+ 1}(z) = z^{\gamma +N+1}$,
\be
%&& \hspace{-1cm} \lim_{z \rightarrow 0} H(z) = 0 , \nonumber \\
%&& \hspace{-1cm} 
H(z) = \frac{ z^{2 \gamma + 2N+2}}{2} - \frac{ z^{4 \gamma + 4N+ 4}}{8} + \dots \,  .
%\frac{ z^{6 \gamma + 6}}{36} + O(z^{6 \gamma + 7}),
\ee
%where the Taylor expansion is the exact one %admitted 
%for ${\rm Arg}(z) < \pi/8$
%but we already have a stronger constraint on the cone $C$, 
%${\rm Arg}(z) < \Theta = \pi/(4 \gamma+4)$. In particular 
%$ \lim_{z \rightarrow 0} H(z) = 0$
%\section{Renormalization}
%In other words t
%The entire function $H(z)$ interpolates between $z^{2 \gamma +2}/2$ at large distances 
%and $\log z^{\gamma+1}$ at short distances.

%%%%%%%%%%%%%%%%%
% UNITARITY EXPLICIT 

\vspace{5mm}
\begin{center}***\end{center}
%\vspace{1mm}

%\begin{center}
%\line(2,0){50}
%\end{center}

We now present a systematic study of the tree-level unitarity \cite{HigherDG}.
A general theory is well defined if  ``tachyons" and ``ghosts" are absent, in which case the corresponding
propagator has only first poles at $k^2 - M^2 =0$ with real masses (no tachyons) and with positive
residues (no ghosts). Therefore, to test the tree-level unitarity of a multidimensional 
super-renormalizable higher derivative gravity we couple the propagator to external conserved 
stress-energy tensors, $T^{\mu \nu}$, and we examine the amplitude at the pole.
When we introduce a general source, the linearized action including the gauge-fixing reads  
\be
\mathcal{L}_{hT} = \frac{1}{2} h^{\mu \nu} \mathcal{O}_{\mu \nu , \rho \sigma} h^{\rho \sigma} 
- g \, h_{\mu \nu} T^{\mu \nu}.
\label{LGM}
\ee
The transition amplitude in momentum space is 
\be
\mathcal{A} = g^2 \, T^{\mu \nu} \, \mathcal{O}^{-1}_{\mu \nu , \rho \sigma} \, T^{\mu \nu} \, ,
\label{ampli1}
\ee
where $g$ is an effective coupling constant. Here, only the projectors $P^{(2)}$ and 
$P^{(0)}$ will give a non zero contribution to the amplitude since the energy tensor 
is conserved. %$k_{\mu} T^{\mu \nu} = 0$.
To make the analysis more explicit, we can expand the sources using the 
following set of independent vectors in the momentum space \cite{HigherDG},
\be
&& k^{\mu} = (k^0, \vec{k}) \, , \,\, \tilde{k}^{\mu} = (k^0, - \vec{k}) \, , \nonumber \\
&& \epsilon^{\mu}_i = (0, \vec{\epsilon}) \, , 
\,\, i =1, \dots , D-2 \, ,
\ee
where $\vec{\epsilon}_i$ are unit vectors orthogonal to each other and to $\vec{k}$.
The symmetric stress-energy tensor reads 
\be
&& T^{\mu\nu} = a k^{\mu} k^{\nu} + b \tilde{k}^{\mu} \tilde{k}^{\nu} + c^{i j} \epsilon_i^{(\mu} \epsilon_j^{\nu)} \nonumber \\
&& + d \, k^{(\mu} \tilde{k}^{\nu)} + e^i k^{(\mu} \epsilon_i^{\nu)} + f^i \tilde{k}^{(\mu} \epsilon_i^{\nu)}.
\ee
The conditions $k_{\mu} T^{\mu \nu} =0$ and $k_{\mu} k_{\nu}T^{\mu \nu} =0$ place constrains 
on the coefficients $a,b,d, e^i, f^i$ \cite{HigherDG}.

Introducing the spin-projectors (\ref{proje}) and the conservation of the stress-energy tensor 
$k_{\mu} T^{\mu \nu} = 0$ in (\ref{ampli1}), the amplitude results 
\be
\mathcal{A} = g^2 \left\{  T_{\mu \nu} T^{\mu \nu} - \frac{T^2}{D-2} \right\} \frac{e^{- H(k^2/\Lambda^2)}}{k^2} \, .
\label{ampli2}
\ee
Clearly, there is only the graviton pole in $k^2 =0$ and the residue at $k^2=0$ is 
\be
{\rm Res} \, \left( \mathcal{A} \right) \big|_{k^2 =0} = g^2 \left[ (c^{ij})^2 - \frac{ (c^{ii})^2}{D-2} \right]|_{k^2 =0}.
\label{residuo}
\ee
For $D>3$ the result above result tells us that ${\rm Res} \, \left( \mathcal{A} \right) \big|_{k^2 =0}>0$,
which means that the theory is unitary.
Instead, for $D=3$ the graviton is not a dynamical degree of freedom and the amplitude is zero.

A first example of this quantum transition 
%Let us look at 
is the interaction of two static point particles. In this case, $T^{\mu}_{ \nu} = {\rm diag}(\rho,0,0,0)$ 
with $\rho = M \, \delta(\vec{x})$ and the amplitude (\ref{ampli2}) simplifies to 
\be
\mathcal{A} = g^2 \, \rho^2 \left( \frac{D-3}{D-2} \right) \, \frac{e^{- H(k^2/\Lambda^2)}}{k^2} \, ,
\label{ampli3}
\ee
which is positive in $D>3$ and zero for $D=3$ since, again, there are no local degrees of freedom in $D=3$.

A second example we want to consider is the light bending in the multidimensional nonlocal gravity.
We consider a static source and a light ray. The amplitude for this process is
\be
\mathcal{A} = g^2 \, T^{\mu \nu} \, \mathcal{O}^{-1}_{\mu \nu , \rho \sigma} \, T_{\rm EM} ^{\mu \nu} \, ,
\label{ampli4}
\ee
where $T^{\mu \nu}$ is the above energy tensor for the static particle and $T^{\mu \nu}_{\rm EM}$
is the traceless electro-magnetic energy tensor. Using the projectors defined in (\ref{proje}) and the 
propagator (\ref{propgauge2}), we obtain 
\be
\mathcal{A} = g^2 T^{00} T_{\rm EM}^{00} \,  \frac{e^{- H(k^2/\Lambda^2)}}{k^2}.
\label{amplib}
\ee
We see that, at low energy ($\Lambda \rightarrow + \infty$), the amplitude (\ref{amplib}) is precisely the amplitude for the interaction between a static source and a light ray in $D$-dimensional linearized Einstein's gravity. 
On the other hand, at high energy the light bending is much smaller in the nonlocal theory
than in the Einstein's one. % the amplitude 

The proposed theory is not unique, but all the freedom present in the action can be read in %a single 
the entire function $V(z)$ or $H(z)$
%of the D'Alembertian operator, 
%$\gamma_{1,2}$ %$H( - \Box/\Lambda^2)$ 
\cite{efimov, Krasnikov, NL1, NL2, NL3, NL4, NL5}.
%($\Lambda$ is a physical mass scale introduced in the classical action). %,
%that in principle can be measured experimentally. %We have a class of sup 
%The function $H( - \Box/\Lambda^2)$ and/or t
The expression ``form factor" for the function $V(z)$ used throughout the paper is not accidental. 
Indeed, It may be read %as ``form factors" 
in analogy with the form factors present in the interaction between a photon and a nucleon. 
Most importantly, in view of this, it   %and then
%The form factor for gravity $V(z)$
%at least in principle, 
can also be measured experimentally. Specifically, the quantities (\ref{ampli3}) and 
(\ref{amplib}), at least in principle, can be measured experimentally to fit $V(z)$. 
The four graviton amplitude is another measurable quantity of the form factor $V(z)$ and it reads 
\be
\mathcal{A}_{\rm 4-grav.} = \mathcal{A}_{\rm 4-grav.} \big(s,t,u; V(s,t,u); \epsilon_{1,2,3,4}  \big) \, ,
\label{A4g}
\ee
where $\epsilon_{1,2,3,4}$ are the four gravitons polarizations and $s,t,u$ the Mandelstam variables.
Since 
$V(z)$ %is not in agreement with 
has to be an entire function, we can falsify the theory by 
comparing the experimental four-gravitons amplitude with (\ref{A4g}). %, we can falsify the theory. 
%if the fit of 

\vspace{5mm}
\begin{center}***\end{center}
%\vspace{-2mm}

%\begin{center}
%\line(2,0){50}
%\end{center}
%\newcommand{\HRule}{\rule{\linewidth}{0.3mm}}
%\HRule \\[2.0cm]

To address the problem of classical singularities
mentioned at the beginning of the paper,  
%the first quantity to calculate in a general modified theory
%of gravity is 
we can start out by calculating the gravitational potential. 
Given the modified propagator (\ref{propgauge2}), the graviton solution of the equations of motion resulting 
from the Lagrangian (\ref{LGM}) with $g= \kappa/2$, is 
\be
&& \hspace{-1.0cm} h_{\mu \nu}(x) = \frac{\kappa}{2} \int d^D x^{\prime} 
\mathcal{O}^{-1}_{\mu \nu, \rho \sigma}(x-x^{\prime}) T^{\rho \sigma} (x^{\prime})  \nonumber \\
&& \hspace{0.2cm} = \frac{\kappa}{2} \int d^D x^\prime \int \frac{d^D k}{(2 \pi)^D} e^{ i k (x - x^\prime)} \nonumber \\
&& \hspace{0.9cm} 
\frac{e^{- H(k^2/\Lambda^2)}}{ k^2} 
\left(T_{\mu \nu} - \frac{\eta_{\mu \nu}}{D-2} T\right).
 \label{hmunuD}
\ee
%where the graviton field is now dimensionless. 
For a static source with energy tensor 
\be
T^{\mu}_{\nu} = {\rm diag}(M \, \delta^{D-1}(\vec{x}), 0, \dots, 0), 
\label{staticSource}
\ee 
the solution of (\ref{hmunuD}) satisfying spherical symmetry
reads 
\be
&& \hspace{-0.8cm} h_{\mu \nu}(r) = - \frac{\kappa M}{2} E_{\mu \nu} \int \frac{d^{D-1} k}{(2 \pi)^{D-1}} \, 
\frac{e^{- i \vec{k} \cdot \vec{x}} \, e^{- H({\vec{k}^2/\Lambda^2)}} }{\vec{k}^2} \nonumber \\
&& 
\hspace{0.3cm}=  - \frac{\kappa M}{2} \, \frac{ \pi^{ \frac{D-3}{2} }}{(2 \pi)^{D-2} } \, 
 \frac{E_{\mu \nu}}{r^{D-3}} \times \nonumber \\
&&  \hspace{-0.6cm} \times \int d p \, p^{D-4} \, e^{-H(p^2/r^2 \Lambda^2)}
 \,_0\tilde{F}_1\left( \frac{D-1}{2} ; - \frac{p^2}{4} \right) \, ,
 \label{hd}
\ee
where 
$\,_0\tilde{F}_1(a;z) = \,_0{F}_1(a;z)/\Gamma(a)$ is the regularized hypergeometric confluent function. 
In (\ref{hd}), we also have introduced 
the variable $p = |\vec{k}| r$ and the matrix
$$E_{\mu \nu} = {\rm diag}\left(\frac{D-3}{D-2}, \frac{1}{(D-2)}, \dots, \frac{1}{D-2} \right).$$
For $r\rightarrow 0$, the entire function $H(z)\approx \log z^{\gamma +N+1}$ and the solution (\ref{hd})
is approximated by 
\be
&& \hspace{-0.4cm}
h_{\mu \nu}(r) \approx - \frac{\kappa M}{2} \, \frac{ \pi^{ \frac{D-3}{2} }}{(2 \pi)^{D-2} } \, 
 E_{\mu \nu} \, \Lambda^{2 \gamma+2 N +2} \, r^{2 \gamma + 2 N +5 -D} \times \nonumber \\
&& \hspace{0.2cm} \times \int d p \,  p^{D - (2 N +4) - 2 \gamma -2} \,_0\tilde{F}_1\left(\frac{D-2}{2} ; - \frac{p^2}{4} \right).
\label{hnearzero}
\ee
The solution (\ref{hnearzero}) is clearly regular near $r \approx 0$ since the exponent of the radial coordinate 
is always positive in any dimension $D$.

The gravitational potential is related to the $h_{00}$ component 
of the graviton field by $\Phi = \kappa h_{00}/2$. Then, %using for a static source %with spherically symmetric 
using (\ref{hd}) %for a static %with energy tensor 
%$T^{\mu}_{\nu} = {\rm diag}(M \, \delta(\vec{x}), 0, \dots, 0)$, 
we get 
\be
&& \hspace{-0.5cm}
%\hspace{0.0cm}
\Phi(r) %\frac{\kappa^2}{4} \frac{D-3}{D-2} 
%\int d^D x^\prime \int \frac{d^D k}{(2 \pi)^D} e^{ i k (x - x^\prime)} \nonumber \\
%&& \hspace{0.8cm} \frac{e^{- H(k^2/\Lambda^2)}}{ k^2} M \int \frac{d^{D-1} \vec{q}}{(2 \pi)^{D-1}} \, 
%e^{i \vec{q} \cdot \vec{x}^{\prime}} \nonumber \\
%&& \hspace{-0.5cm} 
= - \frac{\kappa^2 M }{4} \frac{D-3}{D-2} \int \!\! \frac{d^{D-1} k}{(2 \pi)^{D-1}} 
\frac{e^{- H(\vec{k}^2/\Lambda^2)}   \, e^{ - i \vec{k} \cdot \vec{x}}}{ \vec{k}^2}\nonumber \\
&& \hspace{0.3cm}
= - \frac{G_N M}{r^{D-3}} \, 2 \frac{D-3}{D-2} \, \frac{ \Gamma \left( \frac{D-3}{2}  \right) }{\pi^{\frac{D-3}{2}}} \,
\, \mathbb{F}_D(r) \, ,
\label{pote0} \\
&& \hspace{-0.5cm}
%\ee
%where the function $\mathbb{F}_D(r)$ is defined by 
%\be
 \mathbb{F}_D(r) \equiv \frac{2^{4-D}}{\Gamma \left(\frac{D-3}{2} \right)} \!
\int \! dp %\, p^{D-4} 
\, \frac{e^{-H\left( \frac{p^2}{r^2 \Lambda^2} \right)}}{p^{4 -D}} %\nonumber \\
%&& \hspace{3cm}
\,_0\tilde{F}_1\left( \frac{D-1}{2} ; - \frac{p^2}{4}  \right).
\nonumber %\label{FD}
\ee
\hspace{-0.35cm}
For example, in $D=4$, %we can calculate numerically 
(\ref{pote0}) simplifies to  
%Introducing again the variable 
%$p = |\vec{k}| r$ in spherical coordinates, we get 
\be
\Phi(r) = - \frac{G_N M}{r} \underbrace{\frac{2}{\pi} \int_0^{+\infty} d p \, J_0(p) \, %\frac{\sin(p)}{p} 
e^{- H(p^2/r^2 \Lambda^2)}}_{\mathbb{F}(r)} \, ,
\label{pot4D}
\ee
and (\ref{pot4D}) can be integrated numerically.
In %(\ref{pot4D}) 
this latter, $J_0(p) = {\rm sinc}(p) \equiv \sin(p)/p$ is the Bessel function. 
For small values of the radial coordinate ``$r$" (large values of ``$p$") we get 
\be
\Phi \approx - 2 G_N M \, ({\rm const.}) \, \Lambda^{2 \gamma+2} \, r^{2 \gamma +1},
\label{potes}
\ee
where ${\rm const.} \approx 3 \times 10^7 \, \pi$ for $\Lambda =1$ and $G_N =1$. 
The potential (\ref{potes}) is regular for 
$r\rightarrow 0$ and a
 plot of the exact potential (\ref{pot4D}) for $\gamma =3$ and $M =10$ is given in Fig.\ref{pote}.  
\begin{figure}[ht]
\begin{center}
\hspace{-0.4cm}
\includegraphics[width=4.2cm,angle=0]{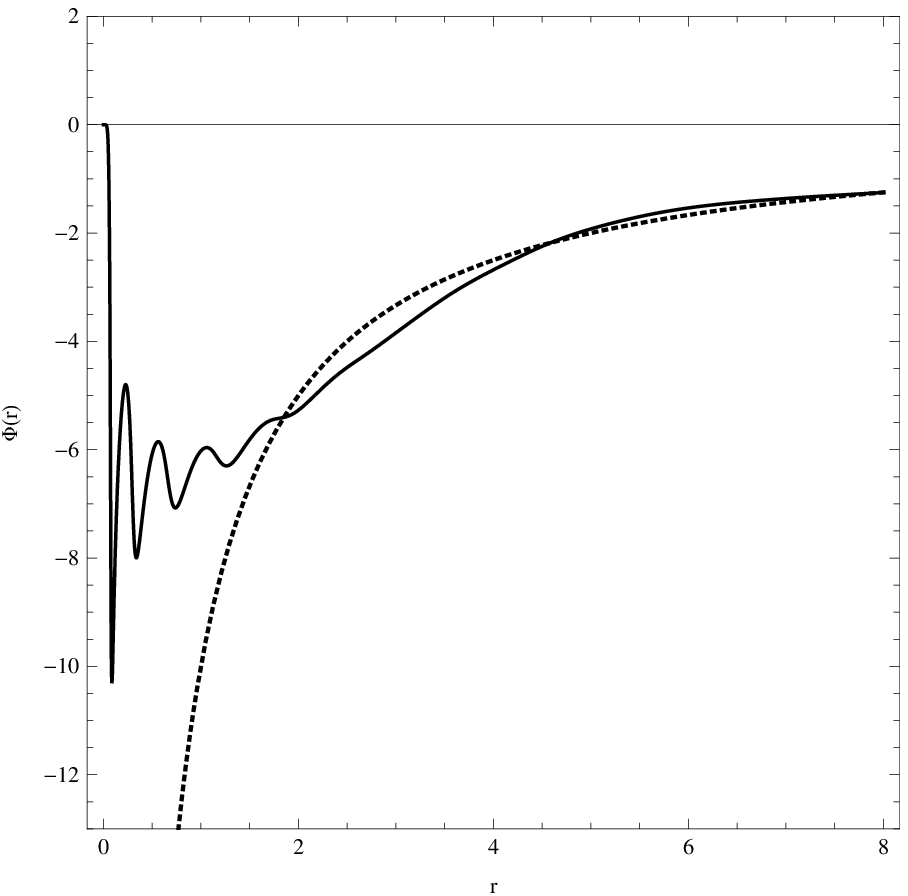}
\includegraphics[width=4.2cm,angle=0]{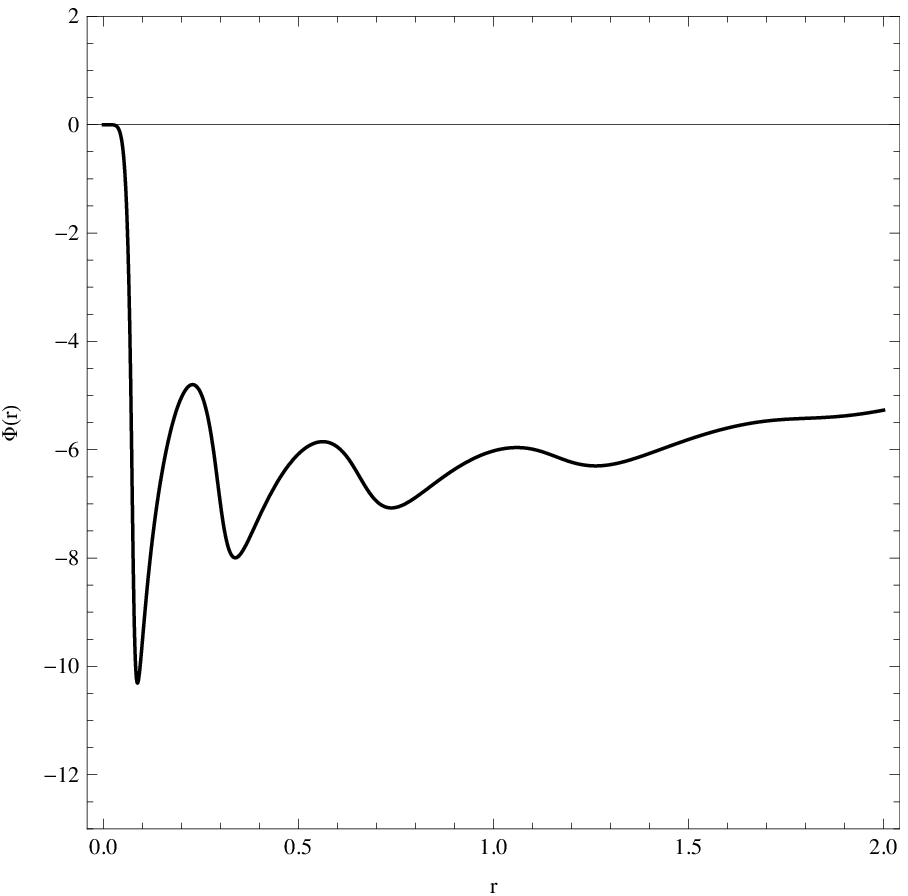}
\caption{\label{pote} Plot of the gravitational potential in $D=4$ for $\gamma =3$ and $M=10$ 
($\Lambda = G_N =1$). }
\end{center}
\end{figure}

Making use of the gravitational potential (\ref{pot4D}), for the sake of simplicity we can study the ``Newtonian cosmology'' 
in $D=4$. 
%for simplicity
%with the modifications coming from (\ref{pot4D}). 
To derive the Friedmann equation 
in Newtonian cosmology, we need the kinetic and potential energy of a test particle and we must 
implement 
energy conservation \cite{NC0, NC}. 
We now consider an observer in a uniform expanding medium of mass density $\rho$. 
Because the Universe is homogeneous and isotropic, we can assume any point to be its center. 
We then identify a particle of mass $m$ at a radial distance $r$. Due to Newton's theorem, the particle only feels a force from the material at smaller values of $r$. The material has mass $M=4 \pi \rho r^3/3$
and the constant total energy of the test particle is $E = T+U$, where $T = m \, \dot{r}^2/2$, 
$U = m \Phi(r)$ and 
$\Phi(r)$ is given in (\ref{pot4D}).
Because the Universe is homogeneous, we apply this argument to any couple of particles,
which allows us to 
introduce comoving coordinates defined by $\vec{r} = a(t) \vec{x}$. 
For the same reason,
the real distance 
$\vec{r}$ is related to the comoving distance $\vec{x}$ by $a(t)$, which is a function of time alone.
%since the homogeneity property of the Universe.
When dividing $E = T+U$ by $a(t)^2 \vec{x}^2$, we get the modified Friedmann equation \cite{BM, BD},
% and relating the energy to the spatial curvature $k$ by 
\be
{\rm H}^2 \equiv \left(\frac{\dot{a}}{a}  \right)^2 = \frac{8 \pi G_N}{3}\,  \rho \, \mathbb{F}(a) - \frac{K}{a^2},
\label{FRW}
\ee
where $K = - 2 E/m \vec{x}^2$ and $\mathbb{F}(a)$ is defined in (\ref{pot4D}). 
To maintain homogeneity, the quantity $E$ must depend on %constant for a given particle does instead 
the comoving  coordinates according to 
%if we look at different separations $\vec{x}$, with 
$E \propto \vec{x}^2$. 
For the same reason,  
we have rescaled $\Lambda^2 \propto1/\vec{x}^2$ obtaining an equation independent from $\vec{x}$
so that homogeneity is respected. Hereon we assume $K=0$. 
As we know the scaling of the gravitational potential for small values 
of ``$r$", we then can get the Friedmann equation near $a(t) \approx 0$.
In this limit we find 
\be
{\rm H}^2 = \frac{8 \pi G_N}{3} \, \rho  \, \frac{ 2 ({\rm const.}) }{\pi} \, a^{2 \gamma +2},
\label{FRW0}
\ee 
where the constant is defined right after (\ref{potes}). It is clear that this Universe is singularity-free, since 
${\rm H} \rightarrow 0$ when $a(t) \rightarrow 0$. Furthermore, if we add the cosmological constant 
$\Lambda_{\rm cc}$ 
to (\ref{FRW}), then the equation (\ref{FRW0}) reads 
\be
{\rm H}^2 = \frac{8 \pi G_N}{3} \, \rho  \, \frac{ 2 ({\rm const.}) }{\pi} \, a^{2 \gamma +2} +\frac{\Lambda_{\rm cc}}{3}.
\label{FRW0Cosmo}
\ee 
This 
shows that the cosmological constant dominates the Universe at high energy whatever 
kind of matter we introduce, either dust or radiation.  
Consequently, the Universe follows a 
natural de Sitter evolution at the Planck scale.
 We can numerically integrate equation (\ref{FRW}) %and the result is given in
 as Fig.\ref{H} shows for both dust matter and radiation.
\begin{figure}[ht]
\begin{center}
\hspace{-0.2cm}
\includegraphics[width=4.1cm,angle=0]{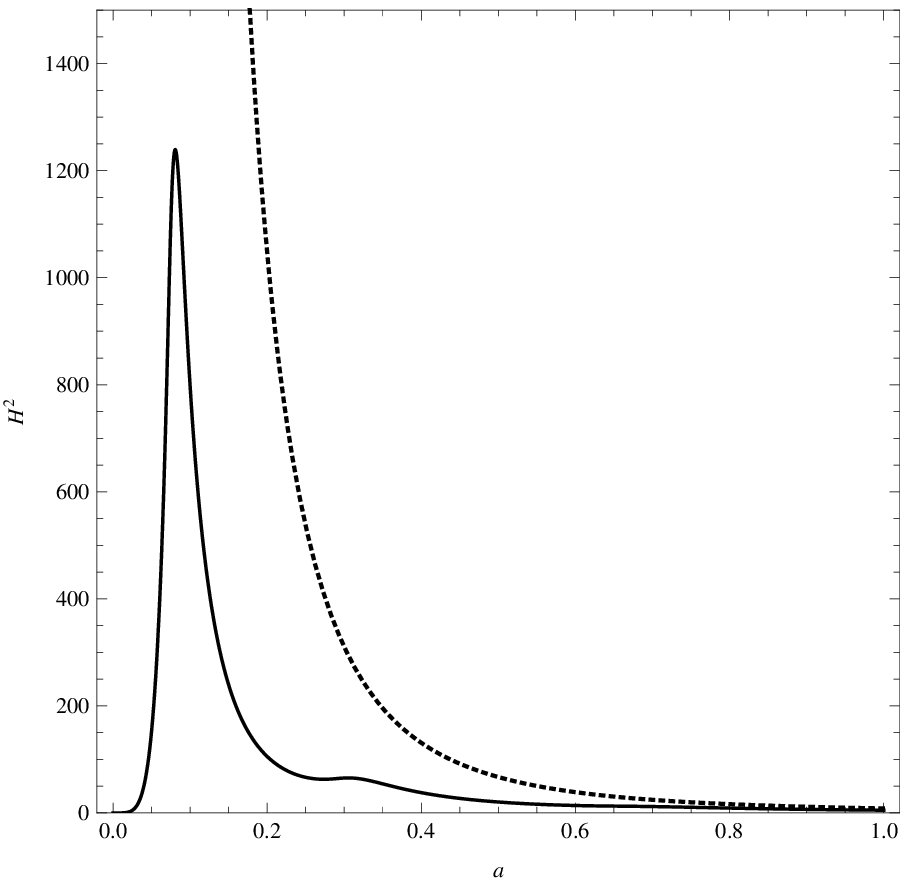}
\includegraphics[width=4.2cm,angle=0]{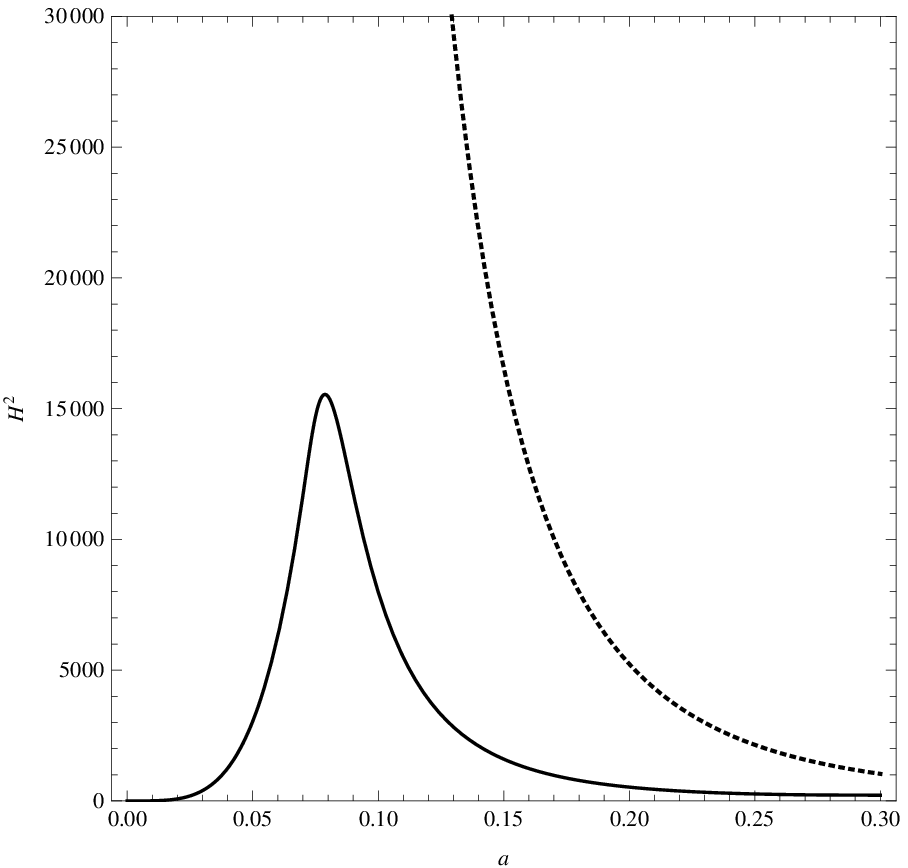}
\caption{\label{H} Plot of ${\rm H}^2(a)$ in $D=4$ for $\gamma =3$ and $\Lambda = G_N =1$ in a universe dominated by dust matter on the left and radiation on the right. The dashed lines account for the classical trajectories.}%and zero cosmological constant.}
\end{center}
\end{figure}

In a $D$-dimensional spacetime, the Newtonian cosmology can be theorized in a way similar to the 
$4$-dimensional case. The Friedmann equation (for $K=0$) 
%coming from 
together with the fluid one, reads %, for zero spatial curvature $K=0$, reads 
\be
&& {\rm H}^2 \equiv \left(\frac{\dot{a}}{a}  \right)^2 = \frac{16 \pi G_N}{(D-2)(D-1)}\,  \rho \, \mathbb{F}_D(a) \, ,\nonumber \\
%- \frac{K}{a^2}
&& \dot{\rho} + (D-1)\left(  \frac{\dot{a}}{a} \right) (\rho + P) =0 \,\,\,\, ({\rm fluid \,\, eq.}) \, ,
\label{FRWD}
\ee
with $\mathbb{F}_D(a)$ as defined in (\ref{pote0}). 
The fluid equation in (\ref{FRWD}) has been obtained using the first law of thermodynamics 
$dE + P\, dV = T \, dS$ ($P$ is the pressure, $T$ the temperature and $S$ the entropy)
and assuming reversible expansion ($dS =0$).
In $D$ dimensions, the state equation for dust matter is $P=0$ and for radiation is 
$\rho = (D-1) P$, so that the fluid equation implies 
\be
\rho_{\rm dust} = \frac{\rho^{0}_{\rm dust}}{a^{D-1}} \,\, , \,\,\,\, \rho_{\rm rad.} = \frac{\rho^{0}_{\rm rad.}}{a^{D}}. 
\ee
When $a \rightarrow 0$, the function 
$\mathbb{F}_D(a) \approx (a \, \Lambda)^{ 2 \gamma +2 N +2}$ %\, ({\rm const.})$ 
and the Friedmann equation simplifies to 
\be
 {\rm H}^2 \equiv \left(\frac{\dot{a}}{a}  \right)^2 \approx \frac{16 \pi G_N}{(D-2)(D-1)}\,  \rho \, 
(a \, \Lambda)^{ 2 \gamma +2 N +2} \, . 
\label{H2D}
%({\rm const.})
\ee
In (\ref{H2D}) ${\rm H}^2$
goes to zero for both radiation and dust matter, which means that 
the $D$-dimensional Newtonian cosmos are singularity-free.

%\vspace{0.2cm}

Consistently with the singularity-free cosmology we have illustrated so far, 
 the black hole solutions turn out to be regular, as 
we are going to demonstrate in $D=4$. 
%%%%%%%%%

Following \cite{Barvi1, Barvi2, Barvi3, Barvi4, Barvi5, Barvi6, Barvi7, Barvi8}, the equations of motion for the above theory (up to square curvature terms)
are 
\be
G_{\mu \nu} + O(R_{\mu \nu}^2) 
+ O(\nabla^2 R_{\rho \sigma})= 8 \pi G_N V(z) %(- \Box/\Lambda^2) 
T_{\mu \nu},
\label{MEE}
\ee
where the argument $z = - \Box_{\Lambda}$ as defined throughout in the paper\footnote{
In general, a differential equation with an infinity number of derivative 
has not a well-defined initial value problem and it needs an infinite number 
of initial conditions. It is shown in \cite{CP1, CP2, CP3} that in a general framework 
each pole of the propagator contributes 
two initial data to the final solution. This is precisely our case because the only pole 
in the bare propagator is the massless graviton and the theory has a well defined 
Cauchy problem.}. 

Since we are going to solve the Einstein equations neglecting curvature square terms, 
then we have to impose the conservation $\nabla^{\mu} (V(z)T_{\mu \nu})=0$
in order for the theory to be compatible with the Bianchi identities. 
Conversely, the exact equations of motion satisfy 
the Bianchi identities because the theory presents general covariance. 
%In some way, t
The condition 
$\nabla^{\mu} (V(z) \, T_{\mu \nu})=0$ compensates the truncation in the modified Einstein equations 
(\ref{MEE}). 

Our main purpose is to solve the field equations by assuming
a static source, which means that the four-velocity field $u^\mu$ has only
a non-vanishing time-like component
%\begin{equation}
$u^\mu\equiv ( u^0 , \vec{0} )$,  % ,\quad
$u^0= (g^{00})^{-1/2}$
\cite{NS1, NS2, NS3, NS4, NS5, NS6, NS7, NS8, NS9, NS10, NS11, NS12, NS13, NS14, NS15,
NS16, NS17, NS18, NS19, NS20}.
%\end{equation}
%Since the spherically symmetric solution, if considered in the domain $0\leqslant r < + \infty$,
%corresponds to a point mass $m$, 
We consider
the component  $T^0\,_0$ of the energy-momentum tensor for a static source of mass $M$ (\ref{staticSource}).
In polar coordinates, 
%is given by 
%\cite{DeBenedictis:2007bm}
%\begin{eqnarray}
 $T^0\,_0= \rho = M \delta(r)/4\pi\, r^2 \label{t00}$ \cite{SpallucciUnp}. 
 %\footnote{
%Usually, in General Relativity textbooks,  %it is customary to introduce 
%the Schwarzschild solution is introduced without mentioning the 
%presence of a point-like source. 
%Once the Einstein equations are solved in the vacuum, the integration constant is determined 
%by matching the solution with the Newtonian field outside a spherically symmetric 
%mass distribution. Definitely, this is not the most straightforward way to expose 
%students %, and not only them, to 
%to one of the most fundamental solutions of the Einstein equations. 
%Moreover, the presence of a curvature singularity in the origin, 
%where from the very beginning a Þnite mass-energy is squeezed into a zero-volume 
%point, is introduced as a shocking, unexpected result. Against this background, 
%we show  %in [36,37,38,39,40] 
%that, once quantum delocalization of the source is 
%accounted, all these flaws disappear. From this follows that for us there is only one physical 
%vacuum solution and this is the Minkowski metric. In other words, the Schwarzschild metric is 
%a vacuum solution with the free integration $m$ equal to zero.
%}.
%because of the spherically symmetric solution, if considered in the domain $0\leqslant r < + \infty$,
%corresponds to a point mass $m$, .
%\end{eqnarray}
The metric of our spacetime is assumed to be given by the usual static, spherically symmetric 
Schwarzschild form
\be
&& ds^2= F(r)dt^2 - \frac{dr^2}{F(r)}-r^2\Omega^2, \nonumber \\
&& F(r)=1-\frac{2 \, G_N \, m(r) }{r}.
\label{metricF}
\ee
The effective energy density and pressures are defined by 
\begin{eqnarray}
 \hspace{-0.5cm} 
%\mathcal{S}^{\mu}{}_{\nu} = 
V(z)T^{\mu}{}_{\nu} %\nonumber \\
=\frac{G^{\mu}\hspace{0.001cm}_{ \nu}}{8 \pi G_N} = 
{\rm Diag}(\rho^{\rm e}, - P_r^{\rm e}, - P_{\bot}^{\rm e}, - P_{\bot}^{\rm e}).
\end{eqnarray}
We temporarily adopt free-falling Cartesian-like coordinates \cite{SpallucciUnp, ModestoMoffatNico} to calculate
the effective energy density, assuming $p_{\gamma +N+1}(z) = z^4$ in (\ref{HD}),
\be
&& \hspace{-0.5cm}  
\rho^{\rm e}( \vec{x}) := V(-\Box_{\Lambda}) T^{0}{}_{0} = 
 M \, V(-\Box_{\Lambda})\, \delta(\vec{x})  \label{rhoeff} \\
 && \hspace{0.5cm} = M  \int \frac{d^3 k}{(2 \pi)^3} \,   e^{-  H(\vec{k}^2/\Lambda^2)  } e^{i \vec{k} \cdot \vec{x}  } \nonumber \\
&&   \hspace{0.5cm} = \frac{ 2 M}{(2 \pi)^2 \, r^3} \int_0^{+\infty} e^{- H(p^2/r^2 \Lambda^2)} p \, \sin(p) \, dp,
\nonumber 
\ee
where $r=|\vec{x}|$ is the radial coordinate. Here we introduced the Fourier-transform 
for the Dirac delta function and 
we also introduced a new dimensionless variable in the momentum space, $p = |\vec{k}| \, r$, where $\vec{k}$ 
is the physical momentum. %From now we assume $p_{\gamma + 1}(z) = 
The energy density distribution defined in (\ref{rhoeff})
respects spherical symmetry. % and is plotted in Fig.\ref{plotrhoeff3}. 
We evaluated numerically the integral in (\ref{rhoeff}) and the resulting
energy density is plotted in Fig.\ref{plotrhoeff3}.
In the low energy limit we can expand $H(z)$ for $z = - \Box/\Lambda^2 \ll 1$ and we can 
 integrate analytically 
(\ref{rhoeff}) %. The result is really involved and only plotted in Fig.\ref{plotrhoeff},
\be
\rho^{\rm e}( r ) 
 = \frac{ 2 M}{(2 \pi)^2 \, r^3} \int_0^{+\infty} e^{-p^{16}/(2 r^{16} \Lambda^{16})} p \, \sin(p) \, dp. 
\label{rhoeff2}
\ee
The result is extremely complex, and its plot is given in Fig.\ref{plotrhoeff3}; however, 
the Taylor expansion near $r \approx 0$ generates a constant leading order 
%\be
$\rho^{\rm e}(r) \propto M \Lambda^3$.%/32 \, 2^{7/16} \Gamma(11/16) \Gamma(7/8) \Gamma(5/4)$.
%\ee
%We evaluated numerically the integral in (\ref{rhoeff}) and the resulting
%energy density is plotted in Fig.\ref{plotrhoeff3}.
\begin{figure}[ht]
\begin{center}
\hspace{-0.4cm}
\includegraphics[width=4.2cm,angle=0]{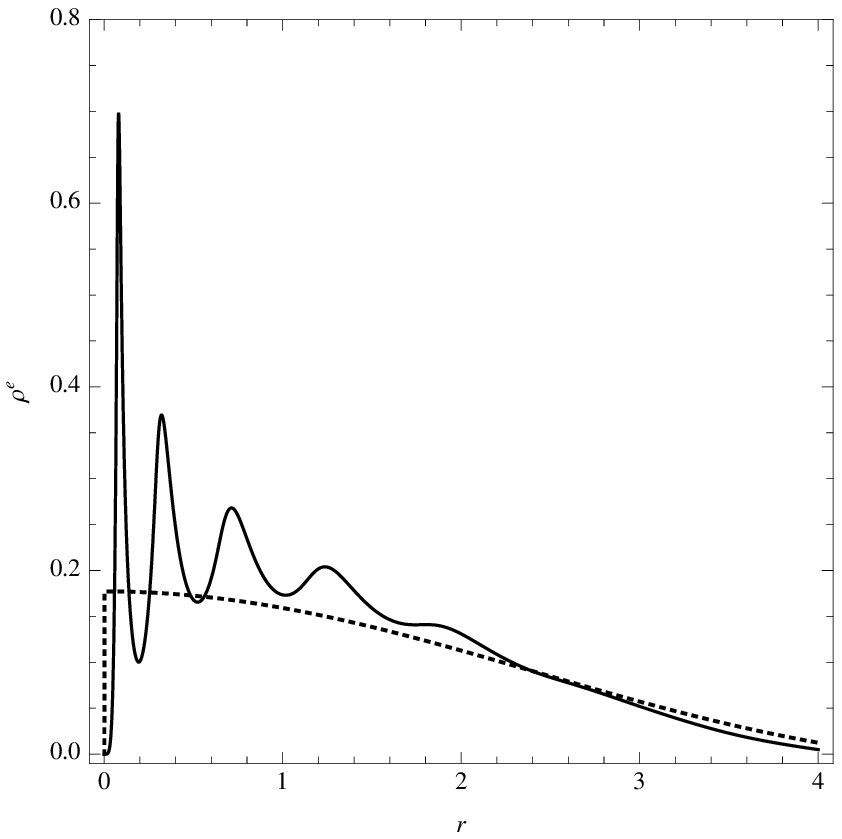}
\includegraphics[width=4.2cm,angle=0]{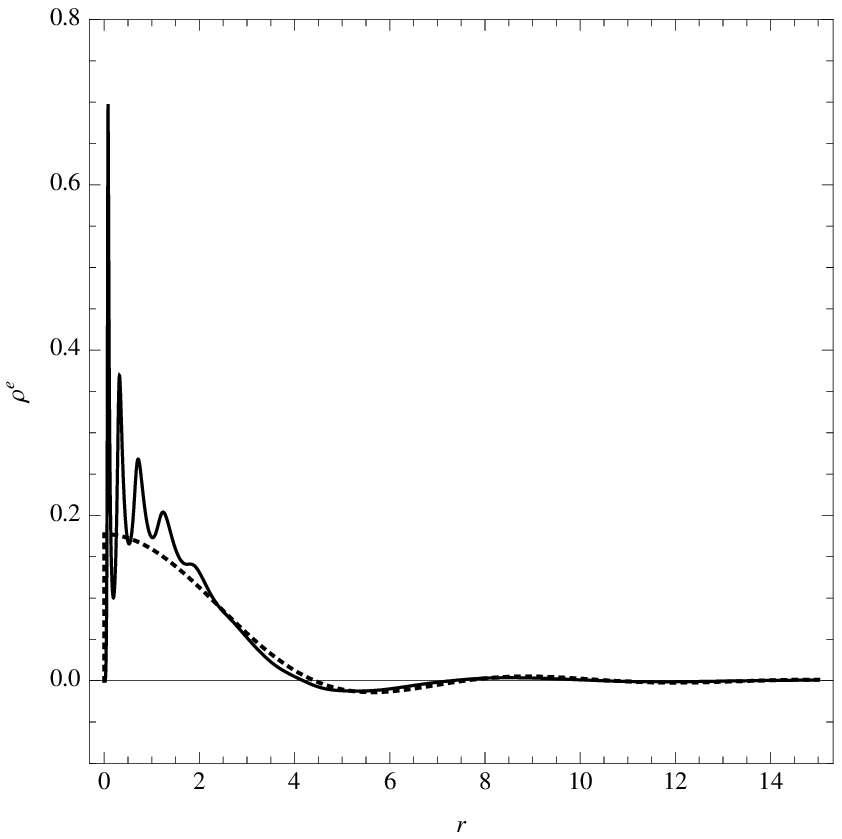}
\caption{\label{plotrhoeff3} Plot of the energy density for $m=10$ in Planck units assuming 
$\Lambda = m_P$. The solid line is a plot of (\ref{rhoeff}) without any approximation. 
The dashed line refers to the 
energy density profile (\ref{rhoeff2}) in the limit $- \Box/\Lambda^2 \ll 1$.}
\end{center}
\end{figure}

The covariant conservation and the additional condition, $g_{00} = - g_{rr}^{-1}$ 
fully specify the form of $V(z)T^{\mu}{}_{\nu}$ and  
the Einstein's equations reads 
\begin{eqnarray}
&& \frac{ d m(r)}{dr} = 4 \pi \rho^{\rm e} \, r^2 ,  \nonumber \\
&& \frac{1}{F} \frac{ d F}{dr} = \frac{2 G_N \, \left(  m(r) + 4 \pi \, P_r^{\rm e} \, r^3\right)}{r (r-2 \, G_N \, m(r))} , \nonumber \\
&& \frac{d P_r^{\rm e}}{d r} = - \frac{1}{2 F}  \frac{ d F}{d r} (\rho^{\rm e} + P_r^{\rm e}) 
+ \frac{2}{r} (P_{\bot}^{\rm e} - P_r^{\rm e} ).
\label{Eineq}
\end{eqnarray}
Because of the complicated energy density profile, this is how the first Einstein equation 
would fit in (\ref{Eineq})
\begin{equation}
m(r) =  4\pi \int_0^r dr' r'^2 \ \rho^{\rm e}(r').
\label{mass}
\end{equation}
 However, the energy density goes to zero at infinity, reproducing the
asymptotic Schwarzschild spacetime with $m(r) \approx M$ (constant). % for $r \rightarrow + \infty$.
On the other hand, it is easy to calculate the energy density profile close to $r \approx 0$
since $H(z) \rightarrow \log z^4$ for $z \rightarrow + \infty$ 
(or $r \rightarrow 0$ in (\ref{rhoeff})). % in the variable $p$
In this regime, $m(r) \propto M \, \Lambda^8 r^8$ and, for a more general monomial 
$p_{\gamma +1}(z) = z^{\gamma +1}$, $m(r) \propto M  (\Lambda \, r)^{2 \gamma +2}$.
The function $F(r)$ in (\ref{metricF}) near to $r\approx 0$ is approximated by  
\be
F(r) \approx 1 - c \, G_N \, M \, \Lambda^{2 \gamma +2} \, r^{2 \gamma +1} ,
\label{core}
\ee
where $c$ is a dimensionless constant. 

We show now that the metric has at least two horizons: an event horizon and a Cauchy horizon.
The metric interpolates two asymptotic flat regions, one at infinity and the
other in $r = 0$, so that we can write the $g_{r r}^{-1} = - F$ component in the following way
\be
F(r) = 1 - \frac{2 \, M {\rm G}(r)}{r}.
\ee
Here ${\rm G}(r) \rightarrow G_N$ for $r\rightarrow  \infty$, ${\rm G}(r) \propto G_N \, r^{2 \gamma +2}$ for 
$r\rightarrow 0$
and ${\rm G}(r)$ does not depend on the mass $M$.
The function $F(r)$ goes to ``$1$" in both limits (for $r \rightarrow + \infty$ and $r \rightarrow 0$).
Since $M$ is a multiplicative constant, %it can always take up a fixed value 
we can always vary it %the mass $M$ 
for a fixed value of
the radial coordinate $r$, such that $F(r)$ becomes negative. %positive everywhere. 
From this it follows that the function 
$F(r)$ must change sign at least twice. The second equation in (\ref{Eineq}) is solved by 
$P_r^{\rm e} = - \rho^{\rm e}$ and the third one defines the transversal pressure once the energy density $\rho^e$ is known.
For the lapse function $F(r)$ in (\ref{core}), 
%For the structure of the metric near $r\approx 0$ given in (\ref{core}) 
we can calculate the Ricci scalar and the Kretschmann invariant
\be
&& \hspace{-0.5cm}  R = c \, G_N \, M \, \Lambda^{2 \gamma +2} \, (2 \gamma +2) (2 \gamma +3) \, r^{2 \gamma -1} ,  \\
&& \hspace{-0.5cm}  R_{\mu\nu\rho\sigma} R^{\mu\nu\rho\sigma}  = \nonumber \\
&& \hspace{-0.5cm} = 4 \, c^2 \, G_N^2 \, M^2 \, \Lambda^{4 \gamma +4} \, \left(4 \gamma ^4+4 \gamma ^3+5 \gamma ^2+4 \gamma +2\right) r^{4 \gamma - 2}. \nonumber 
\ee
By evaluating the above curvature tensors at the origin one finds that they are finite 
for $\gamma > 1/2$ and, in particular, for the minimal super-renormalizable 
theory in $D=4$ with $\gamma \geqslant 3$.

\begin{figure}[ht]
\begin{center}
\hspace{-0.4cm}
\includegraphics[width=4.2cm,angle=0]{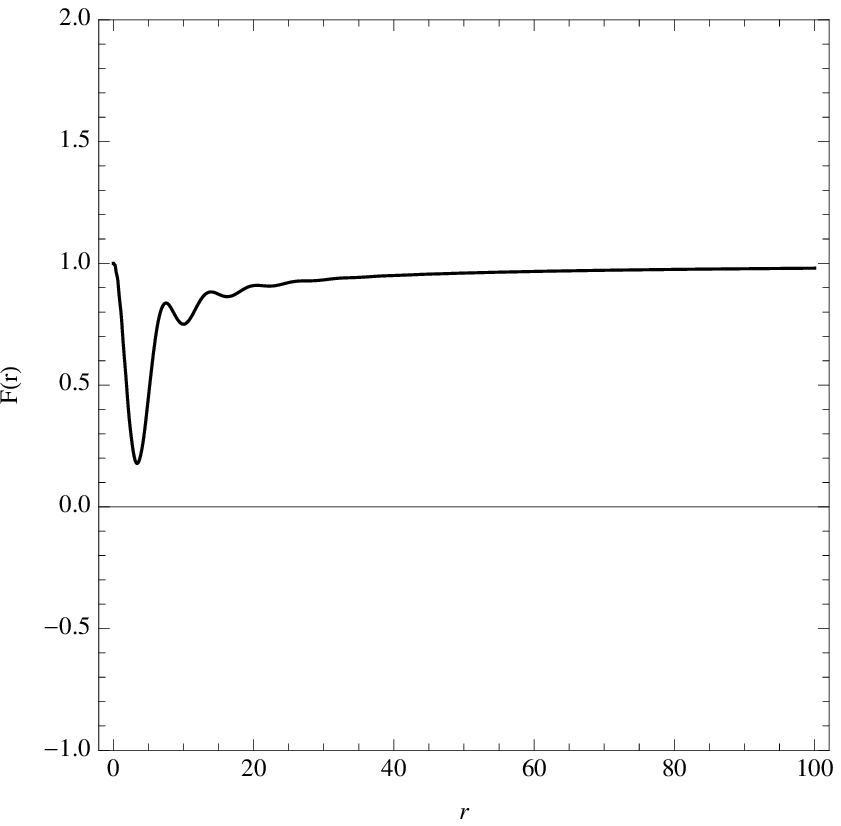}
\includegraphics[width=4.2cm,angle=0]{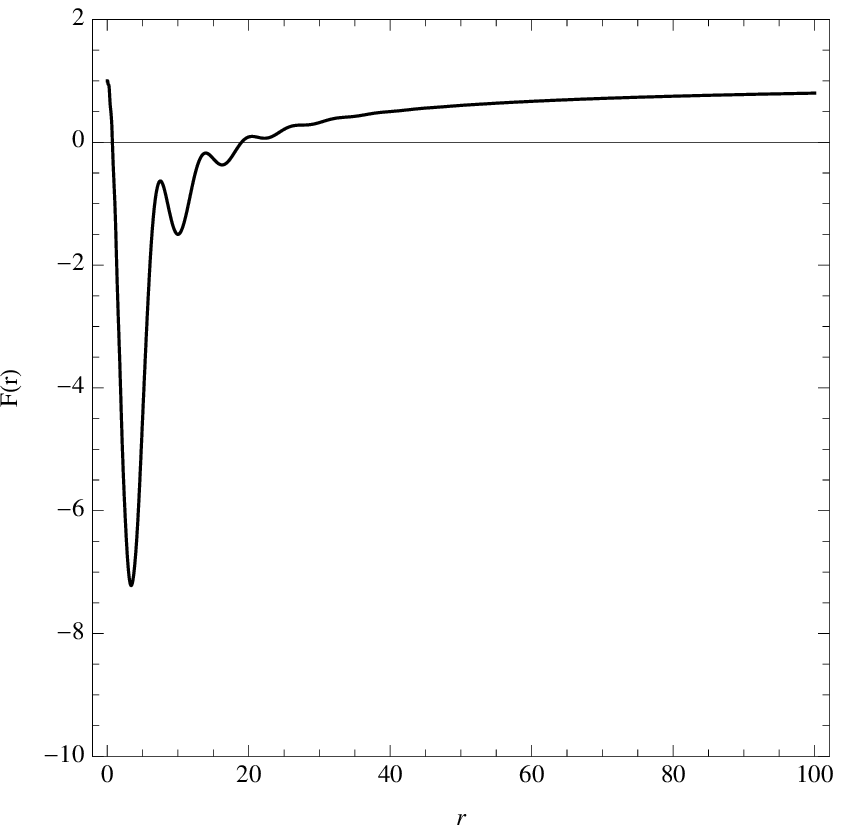}
\caption{\label{Fr2} The first plot shows the function $F(r)$ for the energy profile (\ref{rhoeff2})
and $H(z)$ defined in (\ref{H0}) with the parameter $\gamma =3$.
The ADM mass values are $M=1$ and $M=10$ (in Planck units) for the first and the second plot respectively.
}
%(\ref{ExpH}).}
%\label{ExpH}
\end{center}
\end{figure}
The form factor $V(z)$
is able to tame the curvature singularity of the Schwarzschild solution 
at least for the truncation of the theory here analyzed. 
However, we believe that 
the higher order corrections to the Einstein's equations will not substantially change 
the fundamental remarkable feature of the solutions found in this section \cite{Bob}.

Besides, we can exactly (but only numerically) integrate the modified Einstein equations %of motions 
(\ref{MEE}) for the %two 
energy density defined %respectively 
in (\ref{rhoeff}). % and (\ref{rhoeff2}).
Using the integral form of the mass function (\ref{mass}), we achieve the metric component $F(r)$ 
defined in (\ref{metricF}). 
The numerical results are plotted  
in %Fig.\ref{Fr1} and 
Fig.\ref{Fr2} for different values of the ADM mass $M$. 
The metric function $F(r)$ can intersect no times, twice or more than twice the horizontal axis 
according to the value of the ADM mass $M$. This may enable  
``{multi-horizon black holes}" as an exact solution of the modified equations of motion 
(\ref{MEE}).

 We expect the same features to be maintained in any dimension $D>4$.

%%%%%%%%%%%%%%%%%%%%%%%%%

\vspace{5mm}
\begin{center}***\end{center}
%\vspace{6mm}

%%%%%%%%%%%%%%%%%%%%%%%%%%%
%  EFIMOV 
%%%%%%%%%%%%%%%%%%%%%%%%%%
%\begin{center}%{\leftmargin=1cm}
%\hrulefill
%\dotfill
% \end{center}

%\begin{center}
%\line(2,0){50}
%\end{center}

% \leaders\hbox [1cm] {\hrulefill} \hfill

We can now address a more general class of theories following the Efimov's book 
on non local interactions \cite{efimov}.
Let us consider the propagator in the following general form 
\be
\mathcal{O}^{-1}(z) = \frac{V(z)}{ z \, \Lambda^2}
\label{propgeneral}
\ee
(the notation is rather compatible with the graviton propagator 
(\ref{propgauge2}) and $z := - \Box_{\Lambda}$).

As was shown by Efimov \cite{efimov}%(pag.134)
, the nonlocal field theory is ``unitary" and ``microcausal"
provided that the following properties are satisfied by $V(z)$, 
\begin{enumerate}
\renewcommand{\theenumi}{\Roman{enumi}}
\item
$V(z)$ is an entire analytic function in the complex
$z$-plane and it has a finite order of growth $1/2 \leqslant \rho < + \infty$ i.e. $\exists b>0,c>0$ so that
\be
|V(z)| \leqslant c \, e^{b \, |z|^{\rho}}.
\ee
\item When ${\rm Re}(z) \rightarrow + \infty$ ($k^2 \rightarrow + \infty$), $V(z)$ %is sufficient
decreases quite rapidly. % n a fairly quick way. 
For example, we can consider the following cases. 
\begin{enumerate}
\renewcommand{\labelenumii}{\alph{enumii}.}
    \item $V(z) = O\left(\frac{1}{|z|^a}  \right)$  , $a>\frac{D-2}{2}$.   
For $a = \frac{D-2}{2}$ the theory is not super-renormalizable, but may still be renormalizable.
    \item $\lim_{ {\rm Re}(z) \rightarrow +\infty} |z|^N |V(z)| =0$, $\forall \, N>0$. 
\end{enumerate}
\item  $[V(z)]^{*}= V(z^*)$.
\item $V (0) = 1$.
\item The function $V(z)$ can be non-negative on 
the real axis, i.e. $V(x) \geqslant 0$. %, which always occurs when the 
%nonlocality is introduced into the interaction Lagrangian (see Chap. IV),
%but may be indefinite, if we start from the way
%the introduction of form factors, as outlined above.
\end{enumerate}
Here are some examples of possible functions:
\begin{list}{\labelitemi}{\leftmargin=1cm}
\item[A.] $V_{\rm A}(z) = e^{- z^n}$ for $n \in \mathbb{N}_+$, \\
the weight is $\rho = n < +\infty$,
\item[B.] $V_{\rm B}(z) = \left(\frac{\sin\sqrt{z}}{\sqrt{z}} \right)^{2a}$ \, , % \\
%$d= D_{\rm even}$ in even dimension and \\
%$d = D_{\rm odd} +1$ in odd dimension.
\item[C.] $V_{\rm C}(z) = 2^s \, \Gamma(1+s) \frac{J_s(\sqrt{z})}{(\sqrt{z})^s}$ ($s>0$).
%$d= D_{\rm even}$ in even dimension and \\
%$d = D_{\rm odd} +1$ in odd dimension.
\end{list}
When $V(z) = V_{\rm B}(z)\,\, {\rm or} \,\, V_{\rm C}(z)$, the functions $h_i(z)$ in the action are not entire functions,
so they do no longer meet our minimal requirement. 

A more refined growth 
measure is obtained by defining the order $\rho(\theta_1, \theta_2)$ for $V(z)$ %($z := - \Box_{\Lambda}$)
in the angle 
$\theta_1 \leqslant {\rm arg} \, z \leqslant \theta_2$. 
It is a remarkable property of entire functions that, for appropriate $V(z)$, $\rho(\theta_1, \theta_2)$ may 
range from zero to arbitrarily large values as $\theta_1, \theta_2$ change. 
%This property can be exploited 
%in order to obtain controllable UV behavior. 
The function $V(z)^{-1} = \exp H(z)$ that we have 
introduced and extensively studied in the first part of the paper
%requirement will be that 
%h(z) in (2.1) 
exhibits at most polynomial behavior along the real axis and it is of infinite order 
$\rho = + \infty$ in the full complex plane.  
%regards to the
%definition above.

To expand on the point II.a, we calculate the 
propagator in the coordinate space %, in $D$-dimensions and 
for a general 
form factor $V(z)$. The Fourier transform of (\ref{propgeneral}) reads 
\be
G(x) = \int \frac{d^D k}{(2 \pi)^D} \, \frac{V(k^2 \, \ell^2)}{k^2} \, e^{ i \, k x}  \,  , \,\,\,\, \ell \equiv1/\Lambda \, ,
\label{Gx}
\ee
where we neglected any tensorial structure and we assumed Euclidean signature. 
Changing the existing coordinates into $D$-dimensional spherical ones and integrating 
(\ref{Gx}) in 
the angular variables, we get 
\be
&& \hspace{-1.25cm} G(x) = \frac{\pi^{\frac{D-3}{2}}}{(2 \pi)^{D-1} \,  \Gamma\left( \frac{D-1}{2} \right)} 
\int_0^{+\infty} d u \, \frac{u^{\frac{D-4}{2}} V(u \, \ell^2) }{2} \nonumber \\
&& \sqrt{\pi} \, \Gamma\left( \frac{D-1}{2} \right) \,_0\tilde{F}_1\!\!\left( \frac{D}{2}; - \frac{u \, x^2}{4} \right) ,
\label{propcoord}
%&&  \sqrt{\pi} \, \frac{\Gamma\left( \frac{D-2}{2} \right)}{ \Gamma\left( \frac{D-1}{2} \right)} \,
%\sqrt{\pi} \, \frac{\Gamma\left( \frac{D-3}{2} \right)}{ \Gamma\left( \frac{D-2}{2} \right)} \dots \sqrt{\pi} \, \frac{\Gamma\left( 1 \right)}{ \Gamma\left( \frac{3}{2} \right)} 
\ee
where we have introduced the variable $u = k^2$. 
%and ``$\,_0\tilde{F}_1(a;z) = \,_0{F}_1(a;z)/\Gamma(a)$" is the regularized hypergeometric confluent function. 
From II.a, $V(u \, l^2) = O(1/ u^a)$ for $u \rightarrow + \infty$ and since $\,_0\tilde{F}_1\approx {\rm const.}$
for $x^2 \rightarrow 0$, the propagator in the coincidence limit is finite only for certain values of $a$,
\be \hspace{-0.5cm} 
G(0) \propto \int_0^{+\infty} d u \,  u^{ \frac{D-4}{2} - a} < \infty \,\,\, 
\Longleftrightarrow \,\, a > \frac{D-2}{2}.
\ee
%However the theory can be 
For $D=4$ the two-point function in the coordinate space is 
\be
G(x) = \frac{1}{(2 \pi)^2} \int_0^{+\infty} du \, V(u \, l^2)  \, \frac{J_1( \sqrt{u \, x^2}) }{2 \sqrt{u \, x^2 }}, 
\label{propCoordD4}
\ee
where ``$J_1$" is the Bessel function of the first kind ``$J_n(z)$". 
Using II.a ($u \rightarrow + \infty$) and/or the short distances limit $x^2 \rightarrow 0$, 
the propagator (\ref{propCoordD4}) reads
\be
G(x) = \left\{ \begin{array}{lll} %\hspace{0.1cm}
        O \! \left(   \frac{1}{(x^2)^{1-a}  }\right)
        \vspace{0.2cm}  \,\, {\rm for} \hspace{0.2cm}   0<a<1 \,  ,\\ 
            O \! \left(  \ln(x^2) \right) 
        \vspace{0.2cm}  \,\, {\rm for} \hspace{0.2cm}   a = 1 \,  ,
                  \\
                 O\! \left(  1 \right) \,\, 
        {\rm for} \hspace{0.2cm}  a>1 \,  .
        \end{array} \right.
\label{Gx4Da}
\ee
Only for $a>1$, $G(0) < + \infty$ in the coincidence limit.
This is further evidence that super-renormalizability requires $a > (D-2)/2$.
Later on we will show that we may still have renormalizability for $a = (D-2)/2$
in the case study $D=4$. 

We now move to investigate the general theory (\ref{action}) %renormalizability 
with form factor 
$V_A(z) = \exp(- z^n)$, which satisfies the property II.b., in the entire functions $h_i(z)$ defined in (\ref{hz}). 
%In the momentum space, t
The high energy propagator %omitting the tensorial structure, 
reads %goes as 
\be
 \mathcal{O}^{-1}(k) = \frac{e^{- (k^{2}/\Lambda^{2})^{n}}}{k^{2}}.
 \label{pen}
\ee
The $m$-graviton interaction has the 
same scaling, since it can be written in the following schematic way
%{\small
\be
&&
 {\mathcal L}^{(m)} \sim  h^m \, \Box_{\eta} h \,\,  h_i( - \Box_{\Lambda}) \,\, \Box_{\eta} h \nonumber  \\
&& \hspace{0.7cm}
\rightarrow h^m \, \Box_{\eta} h 
\,  \frac{e^{ \left(- \frac{\Box_{\eta}}{\Lambda^2} \right)^n }}{\Box_{\eta}} \, 
\Box_{\eta} h + \dots  \, , %\nonumber  
\label{intera2}
\ee%}
where $\Box_{\eta} = \eta^{\mu \nu} \partial_{\mu} \partial_{\nu}$. The notation ``$\dots$" indicates other sub-leading interaction terms %with more graviton derivative 
coming from the
covariant D'Alembertian operator.
% is the flat spacetime D'Alembertian.
%
%
Placing an upper bound to the amplitude with 
$L$-loops, we find 
\begin{eqnarray}
&& \hspace{-0.4cm} \mathcal{A}^{(L)} \leqslant \int (d^D k)^L \, \left(\frac{e^{-k^{2n}/\Lambda^{2n}}}{k^2} \right)^I \, 
\left(e^{k^{2n}/\Lambda^{2n}} k^2 \right)^V \nonumber  \\
&& %\hspace{-0.5cm} 
\hspace{0.4cm}
= \int (d k)^{D L} \left(\frac{e^{-k^{2n}/\Lambda^{2n}}}{k^2} \right)^{\!\! I - V} \nonumber \\
&& \hspace{0.4cm}
 = \int (dk)^{D L} \left(\frac{e^{-k^{2n}/\Lambda^{2n}}}{k^2} \right)^{\!\! L-1}  .
\label{diverExp}
\end{eqnarray}
In the last step we used again the topological identity $I = V+L-1$.
The $L$-loops amplitude is UV finite for $L >1$ and it diverges as ``$k^D$" for $L=1$.
%Thus, 
Only 1-loop divergences survive in this theory. Therefore, 
the theory is super-renormalizable and unitary, as well as microcausal as pointed out in 
\cite{efimov, E2, E3, E4, E5, Krasnikov}.

To calculate the gravitational potential for $n=1$, 
%in $D=4$ and $n=1$ in (\ref{pen}), %we can use (\ref{pot4D}) 
%we can 
it suffices to replace
$\exp H(p^2/r^2 \Lambda^2) \rightarrow \exp (p^2/r^2 \Lambda^2)$ within the integral (\ref{hd}). %(\ref{pot4D}). 
The result is:
\be
&&\hspace{-1.1cm}  h_{\mu \nu}(r) = - \frac{\kappa M}{2} \, \frac{ 1 }{4 \, \pi^{\frac{D-1}{2}} \, r^{D-3} } \, 
 E_{\mu \nu} \times \nonumber \\
 && \times  \left [ \Gamma \left( \frac{D-3}{2} \right) -  \Gamma \left( \frac{D-3}{2} ; \frac{r^2 \, \Lambda^2}{4} \right) \right].
\label{hdexp}
\ee
To prove the regularity of the graviton solution, we expand (\ref{hdexp}) near $r = 0$, so that 
we get the following finite  
leading term 
\be
h_{\mu \nu}(0) = - E_{\mu \nu} \, \frac{ \kappa \, M \, 2^{1 -D} \, \Lambda^{D-3}}{(D-3) \, \pi^{\frac{D-1}{2}}}.
\ee
%
%
%The result f
For $D=4$, (\ref{hdexp}) simplifies to 
\be
\Phi(r) = - \frac{ G_N M}{r} \, {\rm Er}\left( \frac{r \, \Lambda}{2} \right).
\ee
The gravitational potential 
is regular in $r=0$ and its value is $\Phi(0) = - G_N M \Lambda/\sqrt{\pi}$. 
For $n>1$, the potential is still regular in $r=0$ and it takes the value $\Phi(0) \propto - G_N M \Lambda$
with a slightly different coefficient. % with coefficient 

In the case $n=1$, we can always solve the equations of motion (\ref{MEE}) for a spherically 
symmetric $D$-dimensional spacetime with metric
\be
ds^2_{D} = F_D(r) dt^2 - \frac{dr^2}{F_D(r)} - r^2 d \Omega_{D-2}, 
\label{MetricD}
\ee
where $d \Omega_{D-2}$ is described in terms of $D-2$ angles. 
The form factor $V(z) = \exp{ -z}$ gives a smearing of the source and the energy density reads 
\be
\rho^{\rm e}(r) = V(z) T^0\,_0 = M \left(\frac{\,\Lambda^2}{4 \pi}\right)^{\frac{D-1}{2}} e^{ -r^2 \, \Lambda^2/4}.
\ee
Integrating the ``$00$" component of the modified Einstein equations (\ref{MEE}), we get the
function $F_D(r)$,
\be
&& F_D(r) = 1- \frac{2 M G_D}{ \Gamma\left(\frac{ D-1}{2} \right) \, r^{D-3}} \, \gamma \left( \frac{D-1}{2} ; \frac{r^2 \Lambda^2}{4} \right) \, \nonumber \\
&& \gamma \left( \frac{D-1}{2} , \frac{r^2 \Lambda^2}{4} \right) \equiv \int_{0}^{r^2 \Lambda^2/4} \frac{dt}{t} 
t^{\frac{D-1}{2}} e^{-t} \,  ,\nonumber \\
&& \Gamma ( (D-1)/2) = \left[ \frac{D-1}{2} - 1 \right]! \,\,\,\, {\rm for} \,\, D \,\, {\rm odd} \, , \nonumber \\
&& \Gamma ( (D-1)/2) = \sqrt{\pi} \left[ \frac{(D-3)!!}{ 2^{(D-2)/2}} \right] \,\,\,\, {\rm for} \,\, D \,\, {\rm even}. 
\label{FD}
\ee
where $[G_D] = M^{2-D}$. 
The other components in (\ref{MEE}) are solved by $\rho^{\rm e} = - P_r^{\rm e}$, % := V(z)T^r\,_r$ 
while the covariant conservation of the effective energy tensor $V(z) T_{\mu \nu}$ determines 
\be 
\hspace{-0.2cm}
T^{{\rm e} \, i}\,_i = - \rho^{\rm e} - (D-2)^{-1} \, r \,  \partial_r \rho^{\rm e} \,\, , \,\,\, i = 1, \dots, D-2 \, .
\ee
The metric has a ``de Sitter core" near the origin $r=0$ where
\be
F_D(r) \approx 1 - \frac{4 M G_D \, \Lambda^{D-1}}{(D-1) 2^{D-2} \pi^{(D-3)/2} } \, r^2,
\ee
%and then the spacetime 
from which descends a singularity-free spacetime. All the other properties of the metric have been extensively 
studied in  \cite{NicoD}.

The repercussions of this study affect
several fields as emerges from previous investigations in
LHC black hole phenomenology \cite{Mureika:2011hg, Nicolini:2011nz}, gauge gravity duality
\cite{Nicolini:2011dp} and early universe cosmology
\cite{NS19, Barvi8}.
Specifically it has been shown that the resulting black
hole tend to emit softer particles on the brane
\cite{Nicolini:2011nz}, a fact which is in marked contrast
with previous results based on classical metrics. 

%In
%addition we recall that the neutrino propagation over
%astrophysical distances can provide a clean experimental
%signature of the theory. Due to nonlocal effects, neutrino
%would fail to oscillate, opening the possibility of
%observations at telescopes such as IceCube and ANTARES.

%\begin{center}%{\leftmargin=1cm}
%\hrulefill
%\dotfill
 %\end{center}
 
 %\begin{center}
%\line(2,0){50}
%\end{center}

\vspace{5mm}
\begin{center}***\end{center}
%\vspace{-2mm}

Another  special theory we wish to explore is defined by the following form factor, 
\be
&& V(z) = e^{-H(z)} \, , \label{modefactor}\\
&& H(z) = \frac{1}{2} \left[ \gamma_E + 
\Gamma \left(0, z^{2N+2}\right) \right] + \log [ z^{N+1} ] \, , 
\nonumber \\
&& {\rm Re}( \, z^{2N+2} ) > 0.
\nonumber 
\ee
This form factor has been achieved from (\ref{HD}) by choosing $\gamma = 0$.
The theory %do not 
satisfies all the properties I - V of the second class of theories here examined. 
%(ii) and (iii) of the first class of theories introduced in the paper.
In particular, the behavior of the entire functions $h_i(z)$ for $|z| \rightarrow + \infty$ is, % is no more the one given in (\ref{tombocond}) but
%instead we have the following, %$h_i(z) \rightarrow {\rm const.}$ 
%\item There exists $\Theta>0$ such that 
\be
&& \lim_{|z|\rightarrow + \infty} |h_i(z)| \rightarrow | z |^{N} \, , \,\,\,\, 
%\,\, , \,\,\,\,  
\label{modecond}
{\rm for} \,\, z  \,\, {\rm in} : \\% the following conical regions 
&& \hspace{-0.2cm} 
C = \{ z \, | \,\, - \Theta < {\rm arg} z < + \Theta \, , \,\,  \pi - \Theta < {\rm arg} z < \pi + \Theta \} , \nonumber \\
&&  \hspace{-0.2cm} 
{\rm for } \,\,\, %0< 
\Theta = \pi/(4 N +4) \, .
%< \pi/2 \, . 
\nonumber 
\ee
Since in even dimensions $N = (D_{\rm even}-4)/2$, the entire functions $h_i(z)$ 
in $D=4$ approache 
a constant for $|z| \rightarrow +\infty$. % \cite{Krasnikov}). 
This theory embodies the quadratic Stelle action in the ultraviolet limit 
but without any ghost pole in the propagator. 
The form factor cross-connects the quadratic action in the infrared
with an equivalent theory in the ultraviolet.

%it is possible to set %(at least at classic level) 

The theory in question 
meets the property II.a for the critical value $a=(D-2)/2$ ($a=1$ in $D=4$) 
\cite{efimov}\footnote{See page 147 and pages 246-252 of the Efimov's study \cite{efimov}.}.
The amplitudes are divergent at each order in the loop expansion and 
the maximal superficial degree of divergence from (\ref{deltaD}) or (\ref{DDX})  
is $\delta = D$ as it occurs in the local theory.  Therefore, the theory ceases to be super-renormalizable, 
but it preserves  renormalizability and unitarity as it can be inferred from (\ref{ampli2}) and (\ref{residuo}) with the entire function $H(z)$ defined in (\ref{modefactor}).

%For a generic form factor the 

The gravitational potential can be obtained integrating (\ref{pot4D}) with
the form factor (\ref{modefactor}). The potential is regular everywhere and $\Phi(r) \approx - 14 \, G_N M \Lambda^2 \, r$
near the point $r=0$.
Because the metric scaling $F(r) \approx 1- ({\rm const.}) \, M \Lambda^2 \, r$ in $r=0$,
black hole solutions are not singularity-free, as proved by the diverging 
curvature invariants \cite{BR, BR2, S2}.

Let us assume that 
the coupling constants in $D=4$ satisfy the following relation (see (\ref{actionRen}))
\be
b_0 (Z_{b_0} - 1) = 3 a_0 (1- Z_{a_0} ) +1
\ee
in the ultraviolet regime. Then the Lagrangian turns out to be 
conformal invariant at high energy \cite{eugenio},
\be
&& {S}_{\rm UV}^{D=4} \propto \int d^4x \sqrt{|g|} \, C_{\mu \nu \rho \sigma} C^{\mu \nu \rho \sigma} \nonumber \\
&&\equiv 2\int d^4x \sqrt{|g|} \left(R_{\mu \nu} R^{\mu \nu} -  \frac{R^2}{3} \right).
\ee
However, the same result can not be achieved 
for the local Stelle's theory, because the relationship between the coupling constants,
for which we get a conformal invariant action, 
is the same one by which the theory loses its renormalizability.  

\begin{figure}[ht]
\begin{center}
\hspace{-0.3cm}
\includegraphics[width=6.0cm,angle=0]{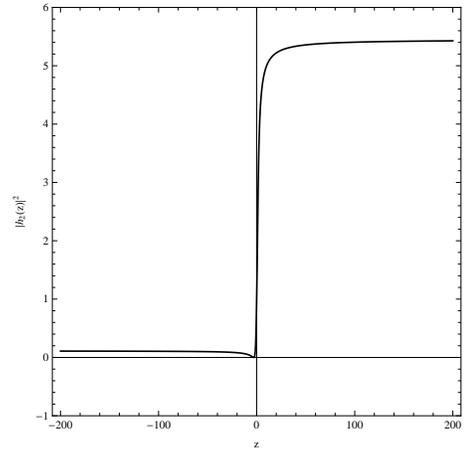}
\caption{\label{h2mode} Plot of $|h_2(z)|^2$ with the form factor defined in (\ref{modefactor}) for $D=4$ and then $N=0$. To draw this plot we have taken $\kappa^2 =2$, $\Lambda =1$ and $\tilde{b}_0 =1$ in (\ref{hz}).
}
%(\ref{ExpH}).}
%\label{ExpH}
\end{center}
\end{figure}
%\begin{figure}[ht]
%\begin{center}
%\hspace{-0.3cm}
%\includegraphics[width=4.2cm,angle=0]{Fr2m1.eps}
%\includegraphics[width=7.5cm,angle=0]{h2zGamma0.eps}
%\includegraphics[width=6.2cm,angle=0]{h2Gamma0ModQuad.eps}
%\caption{\label{h2mode} 3D-Plot of $|h_2(z)|^2$ ($z=r \exp( I \theta)$) with the form factor defined in (\ref{modefactor}) for $D=4$ and then $N=0$. To draw this plot we have taken $\kappa^2 =2$, $\Lambda =1$ and $\tilde{b}_0 =1$ in (\ref{hz}). The radial and angular variables ranges are $r \in[0, 1.7]$ and ${\rm arg} z \in [-\pi , \pi]$. 
%}
%(\ref{ExpH}).}
%\label{ExpH}
%\end{center}
%\end{figure}

%The properties of the function $V(z) = \exp(-H(z))$ above presuppose a region where the weight is zero or in
%other words $V(z)^{-1}$ is asymptotically polynomial  
%%%%%%%%%%%%%%%

%\begin{center}%{\leftmargin=1cm}
%\hrulefill
%\dotfill
% \end{center}

%\begin{center}
%\line(2,0){50}
%\end{center}

\vspace{0mm}
\begin{center}***\end{center}
%\vspace{-2mm}

%%%%%%%%%%%%%%%

%\section{Spectral dimension}
A fundamental quantity that explains ``the spacetime structure" is the spectral dimension (hereafter $d_s$).
This is not only a tool to compare different approaches to quantum gravity, but it is 
actually a device to extract information about the physics within the spacetime. 
It is equivalent knowing either the spectral dimension, the propagator or the gravitational potential.
Schematically,
\be
d_s \,\,\,\, \Longleftrightarrow \,\, {\rm Gravitational \,\,\,\, Potential}.  
\ee
More explicitly, if we know the spectral dimension, then we also know the heat kernel (see below), from which 
we can derive the propagator and the ensuing gravitational potential. 
Clearly, the reverse relationship is true as well.

%On the other hand, the converse 
%is also true.

Here we calculate the spacetime spectral dimension flow from short to long distances
for the three different cases already discussed in the paper,
\begin{list}{\labelitemi}{\leftmargin=1cm}
\item[Form Factor 1.] $V_1(z) = e^{- H(z)} \, ,$
\item[Form Factor 2.] $V_2(z) = e^{-z^n} \,\, , \,\,\,\, n\in \mathbb{N}_+$\,, 
\item[Form Factor 3.] $V_3(z) = e^{\frac{1}{2} \left[ \gamma_E + 
\Gamma \left(0, z^{2N+2} \right) \right] + \log ( z^{N+1} ) }$.
%$z^{N+1} = z^{N+1}
%\left(\frac{\sin\sqrt{z}}{\sqrt{z}} \right)^d$ \,, \\
%$d= D_{\rm even}$ in even dimension and \\
%$d = D_{\rm odd} +1$ in odd dimension.
\end{list}
%showing 
%that the r
As we are going to show,
renormalizability, along with unitarity, implies a spectral dimension 
$d_s <1$ for the the form factor 1, $d_s=0$ for the form factor 2 and $d_s=2$ for the form factor 3.
Let us recall the definition of spectral dimension in quantum gravity.
Such definition is borrowed
from the theory of diffusion processes on fractals \cite{CDT1} and 
adapted to the
quantum gravity context. %\cite{nino,ajl34}. In particular it has been used in
%the Monte Carlo studies mentioned in the Introduction.
%Let us study 
%When studying 
In the Brownian motion of a test particle moving on a $D$-dimensional Riemannian manifold 
$\mathcal M$ with a fixed smooth metric $g^o_{\mu\nu}(x)$,
the probability density for the particle to diffuse from $x'$ to $x$ during the fictitious time %\footnote{
%(this is just a fictitious time since we are probing the spacetime properties, not only the properties of space) %} 
$T$
 is the heat-kernel $K(x,x';T)$. 
This satisfies the heat equation
\begin{eqnarray}
\label{heateq}
\partial_T K(x,x';T)=\Delta_{g^o}^{{\rm eff}} K(x,x';T), 
\end{eqnarray}
where $\Delta_{g^o}^{{\rm eff}}$ denotes %the effective covariant Laplacian that is modified at the quantum gravity  scale. 
%It is 
the usual covariant Laplacian at low energy, which may undergo substantial modifications in the
ultra-violet regime.
In particular, we are
interested in the effective Laplacian at high energy %and in relation to 
on the flat background ($g^o_{\mu \nu} = \eta_{\mu \nu}$) where the graviton propagates.
%\begin{eqnarray}
%\Delta_g\phi\equiv \frac{1}{\sqrt{g}} \,\partial_\mu(\sqrt{g}\,g^{\mu\nu}\,\partial_\nu
%\phi). 
%\end{eqnarray}
The heat-kernel is a matrix element of the operator 
$\exp(T\,\Delta_g)$ acting on the real Hilbert space of %$L^2(\mathcal{M}, \sqrt{|g^o|} \, \text{d}^D x)$, 
%between 
position eigenstates
\begin{eqnarray}
K(x,x';T) =\langle x^{\prime} | \exp(T\,\Delta_{g^o}^{\rm eff}) | x \rangle.
\label{EK}
\end{eqnarray} 
Its trace per volume unit,
\begin{eqnarray}
\label{trace}
%\hspace{-0.35cm} 
&& P(T) \equiv  \frac{\int {d}^Dx\,\sqrt{g^o(x)}\,K(x,x;T)}{V} \nonumber \\
&& \hspace{0.9cm} 
  \equiv   \frac{{\rm Tr}\,
\exp(T\,\Delta_{g^o}^{\rm eff})}{V}
\end{eqnarray}
can be interpreted as an average return probability. Here, $V\equiv\int
d^D x\,\sqrt{g}$ denotes the total volume. It is acknowledged that $P(T)$
possesses an asymptotic expansion for $T\rightarrow 0$ of the form
$P(T)=(4\pi \,T)^{-D/2}\sum_{n=0}^\infty A_n\,T^n$. The coefficients $A_n$ have a geometric meaning, i.e. $A_0$ is the volume of the manifold. % and, if $d=2$, $A_1$ is then proportional to the Euler characteristic. 
Knowing $P(T)$, one can recover the dimensionality of the
manifold $\mathcal{M}$ as the limit for large $T$ of
\begin{eqnarray}
\label{dimform}
d_s \equiv-2\frac{{\partial}\ln P(T)}{{\partial}\ln T}.
\label{defi}
\end{eqnarray}
%If we consider arbitrary fictitious times $T$, this quantity might depend on the scale we are probing. 
This formula %(\ref{defi}) 
defines the fractal dimension we are going to use.

Omitting the tensorial structure in  (\ref{propgauge2}), which does not 
affect the spectral dimension,
we can easily obtain the heat-kernel. % and then the spectral dimension. % of the quantum spacetime. % as experimented from the graviton field.
%In short, in the momentum space the graviton propagator, omitting the tensorial structure that 
%does not affect the 
%spectral dimension, %in the momentum space it 
%reads %has the following behavior, 
%\be
%D(k) \propto \frac{1}{k^2 \, \bar{h}(k^2/\Lambda^2)}.
%\label{propF}
%\ee
We know that the propagator (in the coordinate space) and the heat-kernel are related by 
\cite{manualheat, Vil, Y1, Y2, Y3, Y4, Y5, Y6, Y7}
\be
&& \hspace{-1cm} 
 G(x,x^{\prime}) = \int_0^{+\infty} {\rm d}T\, K(x, x^{\prime}; T)   \label{eatK}\\
&& \hspace{0.3cm} = \int \frac{{d}^D k}{(2 \pi)^D} 
\, e^{i k (x- x^{\prime})} \int_0^{+ \infty}  {d} T \, K(k; T) , \nonumber
\ee
where 
$$G(x, x^{\prime}) = \int \frac{{d}^D k}{(2 \pi)^D} \, e^{i k (x - x^{\prime}) } \, \mathcal{O}^{-1}(k)$$ 
is the Fourier transform 
of (\ref{propgauge2}).
%Given the propagator (\ref{propF}), 
%It is easy to invert 
By inverting (\ref{eatK}) with respect to the heat-kernel in the momentum space, we get 
\be
K(k; T) \propto e^{ - T \, k^2 \, V(k^2/\Lambda^2)^{-1}  } \,  \label{Kk}.
\ee 
%which is the solution of the heat-kernel equation (\ref{heateq}) with the effective operator 
%\be
%\Delta^{\rm eff}_g = \bar{h}( - \Delta_g/\Lambda^2) \, \Delta_g \, , 
%\ee
%which goes like $(- \Delta_g)^{\gamma +1} \, \Delta_g$ at high energy. 
 The necessary trace %we are looking for 
 to calculate the average return probability is 
obtained from the Fourier transform of (\ref{Kk}), %and taking the coincidence limit $x=x^{\prime}$,
\be
K(x,x^{\prime}; T) \propto \int { d}^D k \,  e^{ - T \, k^2 \, V(k^2/\Lambda^2)^{-1}   } \, 
e^{ i k (x - x^{\prime})}.
\ee 
Now we are ready to calculate the average return probability defined in (\ref{trace}) as follows 
\be
P(T) \propto \int {d}^D k \, e^{- T \, k^2 \, V(k^2/\Lambda^2)^{-1} } . %= \frac{\rm const.}{T^{\frac{2}{2 + \gamma}}}
\label{PT}
\ee
We then proceed to calculate explicitly the spectral dimension for the three different 
form factors listed above.

$*$ Form factor 1. 
At high energy, $V(k^2)^{-1} \sim k^{2 \gamma +2 N +2 }$. 
Therefore, we can calculate the integral (\ref{PT}) 
and then the spectral dimension defined in (\ref{dimform}) for small $T$ is 
\be
\hspace{-0.5cm} 
P(T) \propto %\int d^4 k \, e^{- k^2 \, \bar{h}(k^2/\Lambda^2)\, T } 
 T^{- D/(2 \gamma +2 N +4)}  \,\,\,\,\, \Longrightarrow \,\,\,\,\,  d_s = \frac{D}{\gamma + N +2 }.
\ee
Since the parameter $\gamma> D_{\rm even}/2$ or $\gamma > (D_{\rm odd} -1)/2$, 
the spectral dimension 
is $d_s <1 \, \,\, \forall \,\, D$ and 
\be
\lim_{D \rightarrow + \infty} d_s=1 %for $D \rightarrow + \infty,
\ee
applies, which is a ``universal" property of this class of theories.
%manifesting a fractal nature of the spacetime at high energy.
Using the explicit form of the entire function $H(k^2/\Lambda^2)$ given in (\ref{HS}), 
we calculate the spectral dimension at all energy scales as the fictitious time $T$ varies.
%The return probability is 
%\be
%P_g(T) \propto \int {\rm d}^4 k \, e^{- k^2 \, e^{H(k^2/\Lambda^2)}\, T }.
%\label{PH}
%\ee
%We can integrate numerically (\ref{PH}) for $\gamma =3$ 
Integrating numerically (\ref{PT}), we can 
%and to 
plot directly the spectral dimension achieving the graphical result in 
Fig.\ref{SpDim} for $D=4,6,8,10$ and $\gamma =3,4,5,6$, respectively. %\footnote{

\begin{figure}[ht]
\begin{center}
\hspace{-0.3cm}
\includegraphics[width=4.2cm,angle=0]{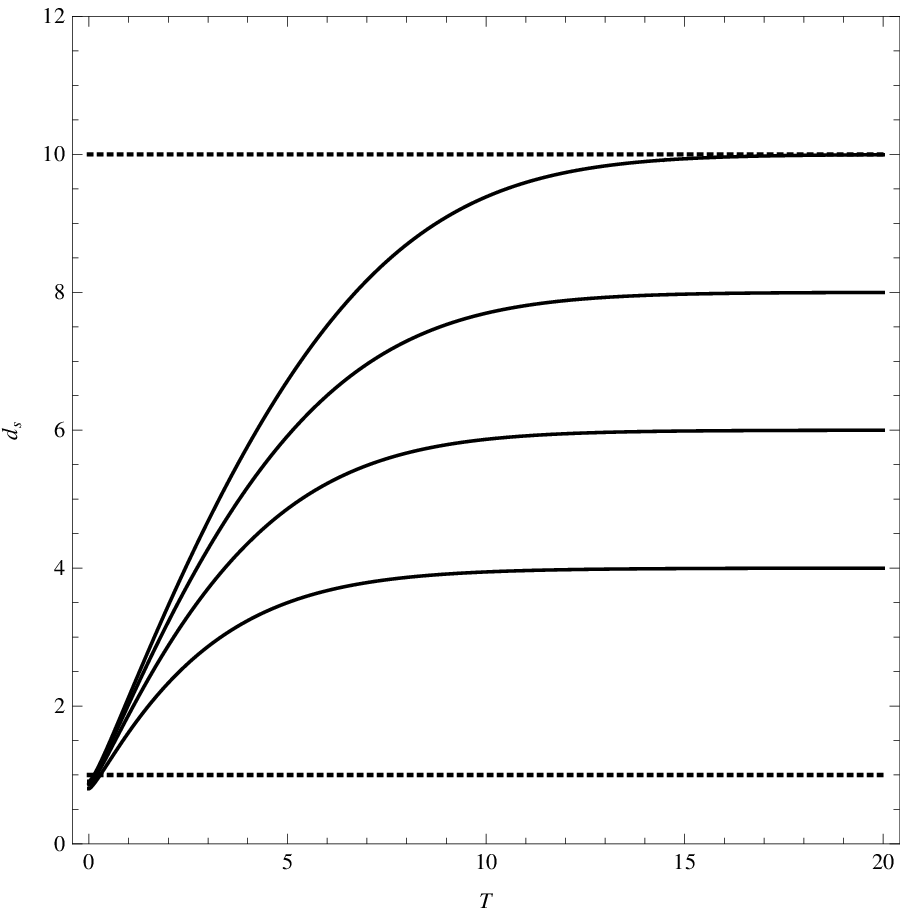}
\includegraphics[width=4.25cm,angle=0]{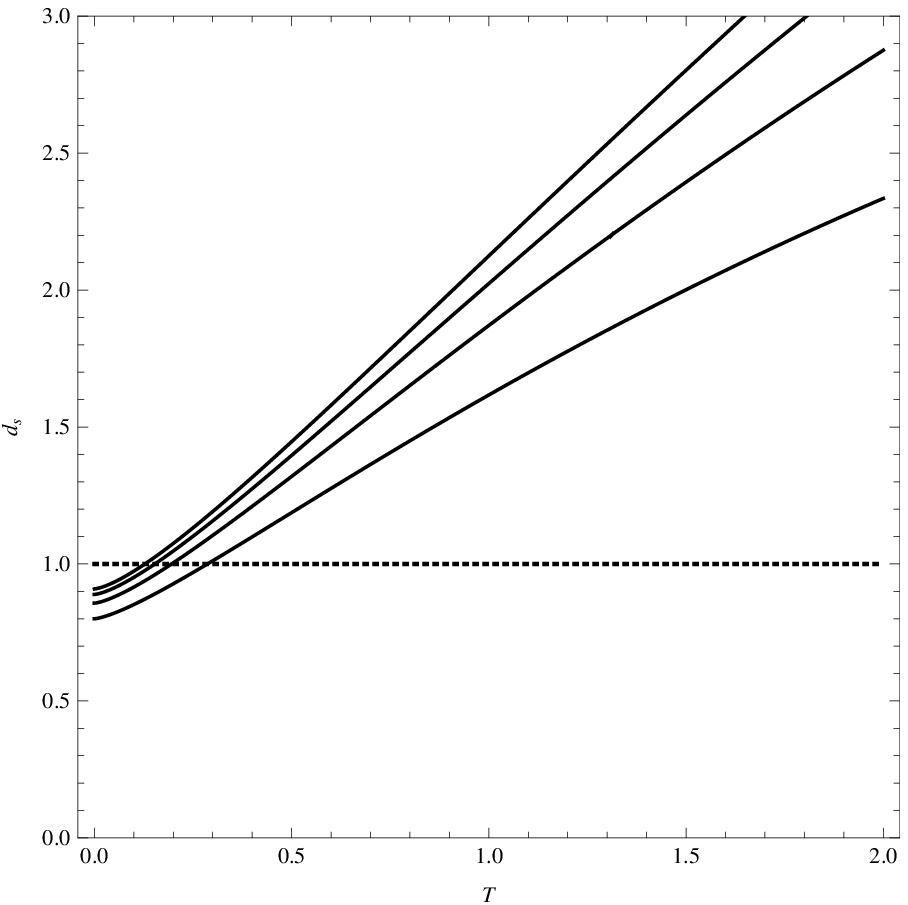}
\caption{\label{SpDim} Plot of the spectral dimension as a function of the fictitious time $T$
for $D=4,6,8,10$ and the minimal values $\gamma=3,4,5,6$ in (\ref{HS}). 
The graph on the right shows that the spectral dimension approximates $d_s =1$ 
when increasing $D$. }
\end{center}
\end{figure}

\begin{figure}[ht]
\begin{center}
\hspace{-0.3cm}
\includegraphics[width=6cm,angle=0]{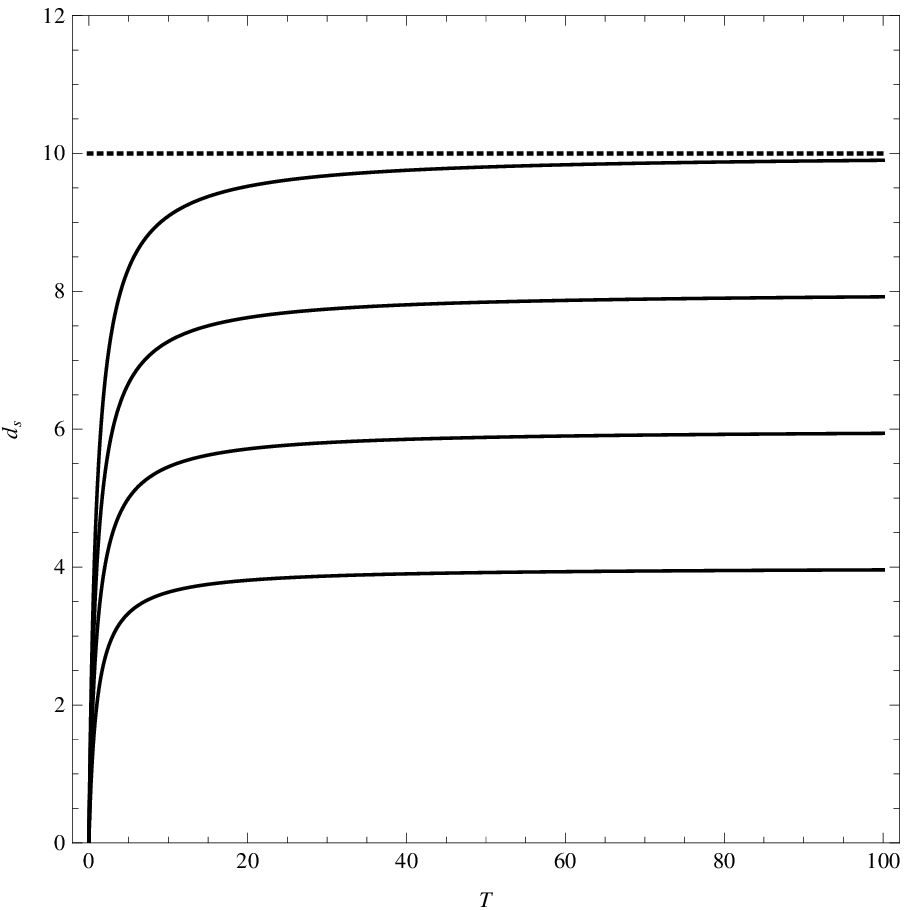}
\caption{\label{SpDimNic} Plot of the spectral dimension as a function of the fictitious time $T$
for $D=4,6,8,10$ and form factor $V(z) = \exp(-z)$. } %in (\ref{HS}). 
%From the picture on the right It is evident that the spectral dimension approximates one 
%with increasing D. }
\end{center}
\end{figure}
\begin{figure}[ht]
\begin{center}
\hspace{-0.3cm}
\includegraphics[width=6cm,angle=0]{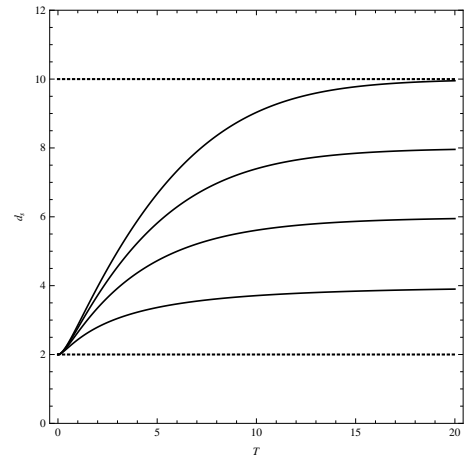}
\caption{\label{SpMode} Plot of the spectral dimension as a function of the fictitious time $T$
for $D=4,6,8,10$ and form factor $V_3(z)$.} % defined in (\ref{HS}). }
%From the picture on the right It is evident that the spectral dimension approximates one 
%with increasing D. }
\end{center}
\end{figure}

$*$ Form Factor 2.  For simplicity we consider the case $n=1$, even though  
 for $n>1$ the result is qualitatively the same.
Given the form factor in the momentum space $V(k^2/\Lambda^2) = \exp(- k^2/\Lambda^2)$, %already 
%introduced in the previous section 
%Another operator we can study is 
%$\exp(-\Box/\Lambda^2)^n$ ($n=1,2$),
the propagator scaling reads 
\be
\mathcal{O}(k)^{-1} \propto \frac{e^{- k^{2}/\Lambda^{2}}}{k^2} \, ,
\label{Dexps}
\ee
%and the spectral dimension goes to zero at high energy.
%In particular, for $n=1$ 
and the heat-kernel can be calculated analytically, % explicitly,
\be
K( x ,x^{\prime} ; T)=\frac{e^{- \frac{\left( x-x^{\prime}  \right)^2}{4 (T + 1/\Lambda^2)
 } }}{\left[  4\pi \left( T+ \Lambda^{-2} \right) \right]^{\frac{D}{2}} } \, ,
\ee
as verifiable %it is easy to verify 
by %going back to the propagator 
(\ref{Dexps}).  
Applying (\ref{dimform}), we find that the spectral dimension is
\begin{eqnarray}
d_s = \frac{T}{T + \Lambda^{-2}} \, D \, ,
\label{piatto}
\end{eqnarray}
which clearly goes to zero for $T \rightarrow 0$ and approaches $d_s =D$ for $T\rightarrow + \infty$. 
A plot of the spectral dimension flow is given in Fig.\ref{SpDimNic} for $D=4,6,8,10$.

$*$ Form factor 3. In this case the spectral dimension in the ultraviolet regime is $d_s =2$, $\forall \, D$.
A plot of the spectral dimension flow is given in Fig.\ref{SpMode} for $D=4,6,8,10$.

It is remarkable to note that for all the three classes of theories parametrized by $(\gamma, n, D)$ and studied in this section, we always find an accumulation point 
 for the spectral dimension in the ultraviolet regime.
 In other words, once perturbative renormalizability and unitarity have set the form factors, the spectral dimension 
 in the ultraviolet regime flows to the same ``critical point" or 
 ``accumulation point" %short distance value 
 independently from the topological dimension $D$.
 From this evidence, we can infer that any consistent theory of quantum gravity 
 must satisfy the following fractal property, %have spectral dimension $d_s \leqslant 2$ in the ultraviolet regime,
% independently from the topological dimension $D$,
\be
d_s \leqslant 2  \, \,\,\,\, \forall \, D \,\,\,\, \rm{in \,\, the \,\, ultraviolet \,\, regime}.
\ee

\vspace{0cm}
%\begin{center}***\end{center}
\section*{CONCLUDING REMARKS} %SUMMARY \& CONCLUSIONS}
%Concluding Remarks 
%\vspace{-2mm}

This study is a synthesis of concepts coming from nonlocal quantum field theory \cite{efimov}, 
particle physics, 
general relativity and string field theory \cite{W1, W2, W3, CSFT, Bis1, Bis2}.
%In this paper we introduce a conservative 
In this article we suggested ways to quantize gravity, %perturbativelly and in the flat Minkowski background
relying on the perturbative approach that has been so successful
%theoretically and experimentally 
for the other fundamental forces. 
% dependent framework.
%
%go now to summarize the main results of this paper.
%In this paper w
We introduced a nonlocal extension of the higher-derivative gravity, which is perturbatively 
renormalizable and unitary in any dimension $D$. %nonlocal 
The four-dimensional theory is easily obtained from the Stelle theory \cite{Stelle} %higher-derivative 
%gravity %that involves an infinite number of derivative terms %. quadratic in the curvature tensor. 
%All the extra operators added to the classical Einstein-Hilbert action are 
by introducing in the action two entire functions, a.k.a. ``form factors",
between the Ricci tensor square and the Ricci scalar square, % as shown below, 
\be
R^2 \, &\rightarrow& \, R \, h_0( - \Box/\Lambda^2) \,R \, , \nonumber \\
 R_{\mu \nu} R^{\mu \nu} \,
&\rightarrow& \,
R_{\mu\nu} \, h_2( - \Box/\Lambda^2) \, R^{\mu \nu}. \label{deca}%\nonumber 
\ee
In the multidimensional spacetime we preserved the two %operators 
``delocalization-operators" as in (\ref{deca}) %defined 
%two entire functions 
and we implemented a finite number of local operators
required (and/or generated) by the quantum consistency of the theory. 
These %most general theory will contain all the possible 
local operators ${\mathcal O}_{2n}(\partial g)$ contain $2n$-derivatives of the metric 
tensor up to 
the mass dimension $[{\mathcal O}_{2n}(\partial g)] \leqslant M^{D}$.
The action may also present other irrelevant operators, whose couplings constants %to these operators 
have negative mass dimension. 
The full action reads 
%%%
\be 
&& \hspace{-1.1cm} 
S = \int d^D x \sqrt{|g|} \Big[2\, \kappa^{-2} \, R + \bar{\lambda} %+ O(R \, (-\Box_{\Lambda})^{N+1} \, R) 
\label{actionF} \\
&& \hspace{-1.1cm}
+ \sum_{n=0}^{N} \Big( 
a_n \, R \, (-\Box_{\Lambda})^n \, R  + 
b_n \, R_{\mu \nu} \, (-\Box_{\Lambda})^n \, R^{\mu \nu}  
%a_n \, R \, (-\Box_{\Lambda})^n \, R 
\Big) 
\nonumber \\
&& \hspace{-1.1cm} 
%+ R_{\mu \nu} \, h_2( - \Box_{\Lambda}) \, R^{\mu \nu} +
+ R  \, h_0( - \Box_{\Lambda}) \, R +
R_{\mu \nu} \, h_2( - \Box_{\Lambda}) \, R^{\mu \nu}  \Big]
 \, \nonumber \\
&& \hspace{-1.1cm}
+ %\hspace{-0.4cm}
\underbrace{ O(R^3) \dots + R^{N+2}}_{\rm Finite \,\, number \,\, of \,\, terms } 
 %\hspace{-0.35cm} 
 + \,
 \underbrace{O(R^{N+3}) + O(R \, (-\Box_{\Lambda})^{N+1} \, R) }_{\rm Non \,\, renormalized \,\, operators} .\hspace{-0.8cm}
%\hspace{0.7cm} \Big] 
 \nonumber 
\ee 
The main reason for introducing the entire functions $h_2(z)$ and $h_0(z)$ is to avoid   
ghosts (or rather the poltergeists: states of negative norm) 
%like the one in the four-dimensional Stelle's theory
and any other new pole in the graviton propagator.
%The theory is indeed ghost-free since the two entire functions 
%have the property to generalize the local higher-derivative gravity 
%in a non-local higher-derivative gravity without 
%introducing new poles in the propagator. 
%By expanding the form factors to the lowest order in a mass scale we introduce,
%the local high derivative theory 
%is recovered. 
%Any truncation of the entire functions gives rise to  
%the unitarity violation and it is only by keeping all the infinite series 
%that we overcome similar issues.  
The unitarity requirement implies the following entire functions,
%If we want a ghost-free theory, the two entire functions, $h_2(z),h_0(z)$, are constrained to 
%the following form,
\be
&& \hspace{-0.5cm} h_2(z) = \frac{ V(z)^{-1} -1 - \frac{\kappa^2 \Lambda^2}{2} \, z 
 \sum_{n=0}^N \tilde{b}_n \, z^n  
}{\frac{\kappa^2 \Lambda^2}{2} \, z}    \nonumber \\
&& \hspace{0.8cm} - O(R \, (-\Box_{\Lambda})^{N+1} \, R) )
\, , \nonumber \\
&& \hspace{-0.5cm} h_0(z) = - \frac{V(z)^{-1} -1 + \kappa^2 \Lambda^2 \, z 
 \sum_{n=0}^N \tilde{a}_n \, z^n}{
\kappa^2 \Lambda^2 \, z} 
\nonumber \\
&& \hspace{0.8cm} - O(R \, (-\Box_{\Lambda})^{N+1} \, R) )
\, . 
\label{hzF}
\ee
The first set of operators in the last line of (\ref{actionF}) is subject to renormalization at quantum
level, whereas the second set %is not renormalized and 
remains classical %under quantization 
as it can be proved 
by power-counting arguments.
%The presence of the last set of operators fix the shape of the entire functions $h_2(z),h_0(z)$
%by unitarity. 
%Since 
Clearly, the non-renormalized operators 
$O(R \, (-\Box_{\Lambda})^{N+1} \, R) )$
 in (\ref{hzF}) %in the definitions of $h_2(z),h_0(z)$ 
%are non 
%renormalized, %at quantum level, 
can be eliminated in both the action and the entire functions. % from the beginning.

The form factors $V(z)^{-1}$ studied in this paper can essentially show two possible high energy 
behaviors, either 
polynomial or exponential.
%However, i
In the first case, the operators 
%If $V(z)^{-1}$ has a 
%polynomial asymptotic behavior, 
%then it is affected by the presence of the operators 
$O(R^{N+3})$ may affect %the asymptotic polynomial degree of $V(z)^{-1}$,
%thus invalidating
the renormalizability of the theory, 
%$O(R \, (-\Box_{\Lambda})^{N+1} \, R)$.
%In other words, if such operators are present in the theory,
%then the renormalizability 
therefore the polynomial asymptotic degree of $V(z)^{-1}$ has to be increased. 
%However, it is sufficient one experimental measure to fix at the same time $V(z)$ 
%and the local classical terms compatibly with unitarity.
In the second case, the same local operators do not thwart the renormalizability of the theory at all.
%if the form factor 
%$V(z)$ decays exponentially in the ultraviolet regime then the theory is not affect at all by 
%the irrelevant local operators. 

Let us gather here all the quantities to be measured to define the theory,
%The quantity to measure experimentally are: 
\be
&& \hspace{-0.3cm}
(a_n, b_n) \, , \,\, 0 \leqslant n \leqslant N \, , \label{couplings} \\
&&  \hspace{-0.3cm}
O(c_n \, R^n) \, , \,\, 3 \leqslant n \leqslant N+2 \,\, ({\mbox {\rm local relevant operators} }) , \nonumber \\
&&  \hspace{-0.3cm}
V(z) \,  , \,\, {\mbox {\rm non-local form factor} } \, , \nonumber \\
&&  \hspace{-0.3cm}
O(d_n \, R^n) \,  , \,\, n \geqslant N+3  \,\, ({\mbox {\rm local irrelevant operators} }) . \nonumber 
\ee
%
%%%
As pointed out throughout the paper, 
the irrelevant operators can always be introduced in any physical 
theory as long as they do not invalidate its unitarity and renormalizability. % properties of the action.
The most important step we need to take is to assess
%is instead 
whether 
the physical measurable quantities are affected or not  
%
%, but if all the physical measurable quantities are not affected 
by such irrelevant operators in (\ref{couplings}). 
If we can assume that no physical quantity is susceptible to such operators, we can then empirically infer 
that the coupling constants equal ``zero".

The question lingers whether 
%Is 
the form factor 
$V(z)$ is measurable or not. % with a finite number of measures?
In principle we can treat $V(z)$ as %to the measure of  
one of the form factors for the scattering of the nucleus by electrons.  
In a gravitational theory, such measure represents the graviton scattering amplitude, 
as well as the modifications to the gravitational potential or the light-bending.
The four-gravitons amplitude will have the general structure 
\be
\mathcal{A}_{\rm 4g.} = \mathcal{A}_{\rm 4g.} \big(s,t,u; V(s,t,u); \epsilon_{1,2,3,4}  \big) \, ,
\label{A4gF}
\ee
where $\epsilon_{1,2,3,4}$ are the four gravitons polarizations and $s,t,u$ the Mandelstam variables.
Since 
$V(z)$ %is not in agreement with 
has to be an entire function, we can falsify the theory by 
comparing the experimental four-gravitons amplitude with the
theoretical prediction (\ref{A4gF}). 
A final remark about the tree-level unitarity of the theory has to be put forward at this point.
Although we consider this issue still %fully 
open \cite{d2, d3, d4, giddi}, we must also acknowledge that, 
%From a physical point of view, 
at high energy, the total cross section of a nonlocal interacted theory 
must not exceed that of a local one. 
Intuitively, the reason is that ``nonlocal particles" must 
manifest the transparency property at high energy because of their non-zero size. 
For example, if the amplitude in the momentum space grows exponentially 
\be
\mathcal{A} \approx e^{(k^2/\Lambda^2)^{\rho}} \,\, , \,\,\, \rho \geqslant 1 \, ,
\ee
the total cross-section will satisfy the following upper bound \cite{E5},
\be
\sigma_{\rm tot}(s) \leqslant {\rm const.} \, s^{\rho -1} \, \log^2(s) \, ,
\ee
which grows logarithmically in the ultraviolet regime if $\rho =1$.
%
%The theory is predictive and T

%What about some possible choices of the form factor $V(z)$? 

%In the paper we considered the power counting renormalizability of three class of theories.
In this paper, we studied the following three classes of form factors $V(z)$, 
\begin{list}{\labelitemi}{\leftmargin=1cm}
\item[Form Factor 1.] $ \,\, V_1(z) = e^{- H_{\gamma}(z)}$\, , 
\item[Form Factor 2.] $ \,\, V_2(z) = e^{-z^n} \,\, , \,\,\,\, n\in \mathbb{N}_+ $\, , 
\item[Form Factor 3.] $ \,\, V_3(z) = e^{- H_{\gamma=0}(z)}$ \, ,
\end{list}
where 
\be \hspace{-0.4cm}
H_{\gamma}(z) = \frac{1}{2} \left[ \gamma_E + \Gamma \left(0, z^{2 \gamma +2 N+2} \right) \right] + \log (z^{\gamma +N+1}). \label{HDc}
\ee
We systematically showed the 
power-counting renormalizability and the tree-level unitarity. 
The theories defined by the form factors $V_1(z)$ and $V_2(z)$ result 
to be renormalizable at one loop and finite from two loops upward.
More precisely, the theories turn out to be super-renormalizable because the covariant counter-terms 
have less derivatives then the classical action and the coefficients of the terms with more
derivatives do not need any kind of infinity 
renormalization as synthesized in the first part of this section.Ê
However, we argue that a supersymmetric extension of the theory \cite{Sugra} can make it finite 
at one loop as well.
For the third choice $V_3(z)$, the theory is merely renormalizable, and %on the contrary of the local theory
no other pole beyond the graviton one appears in the propagator. 

%%%%%%%%%%%%%%%
%At the classical level w
We solved the linearized equations of motion and 
we proved that the gravitation potential is regular in $r = 0$ for all the choices of form factors 
%considered in the paper 
compatible with renormalizability and unitarity. 
We also included Black hole spherical symmetric 
solutions %for two particular choices of the form factors and %are also studied 
omitting higher curvature corrections to the equation of motions. 
For two out of three form factors ($V_1(z)$ and $V_2(z)$) the solutions are regular and the classical singularity
is replaced by a ``de Sitter-like core" in $r=0$. For the third choice $V_3(z)$, black holes are still singular,
although the divergence is attenuated.
%Some preliminary results at 

%For one particular %asymptotically polynomial 
%choice of the form factors
For $V_1(z)$, we proved that the ``Newtonian cosmology" is singularity-free
in any dimension $D$ 
and the Universe %naturally has 
spontaneously follows 
a de Sitter evolution at the ``Planck scale" for any matter content (either dust or radiation),
since the cosmological constant dominates the effective energy tensor at high energy.
In a $D$-dimensional spacetime the modified Friedmann equation (for $K=0$ and $\Lambda_{cc} =0$) reads 
\be
{\rm H}^2 %\equiv \left(\frac{\dot{a}}{a}  \right)^2 
= \frac{16 \pi G_N}{(D-2)(D-1)}\,  \rho \, \mathbb{F}_D(a) \, ,
\ee
where $\mathbb{F}_D(a) \approx (a \, \Lambda)^{ 2 \gamma +2 N +2}$ for $a \approx 0$. 
%We conclude the article considering 
%Black hole spherical symmetric 
%solutions for some particular choices of the form factors and %are also studied 
%omitting high curvature corrections to the equation of motions. The solutions are regular and the classical singularity
%is replaced by a ``de Sitter-like core" in $r=0$. %Black holes may show a ``multi-horizon" structure 
%depending on the value of the ADM mass.

Finally, we have provided an extensive analysis of the spectral dimension  
for any  
$D$ and for the three classes of theories. 
In the ultraviolet regime, the spectral dimension takes on different values
for the three cases: 
%less than or equal to ``$1$" 
%for the first case, ``$0$" for the second one and ``$2$" for the third one.
%Schematically: 
\begin{list}{\labelitemi}{\leftmargin=1cm}
\item[$V_1(z)$] \,\, $\Longrightarrow \,\,\,\,  d_s \lesssim 1$\, , 
\item[$V_2(z)$]  \,\, $\Longrightarrow \,\,\,\, n \in \mathbb{N}_+ \,\,, \,\,\, d_s = 0$\,, 
\item[$V_3(z)$]  \,\,  $\Longrightarrow \,\,\,\, d_s = 2\, .$ \,
\end{list}
where $H_{\gamma}(z)$ is defined in (\ref{HDc}). 
Once the class of theories compatible with renormalizability and unitarity is defined, 
the spectral dimension has the same short-distance ``critical value" or ``accumulation point" for any 
value of the topological dimension $D$. This is a ``universal" property of the theories here studied.

We would like to conclude this section by identifying some similarities between the second class of 
super-renormalizable theories and ``string field theory".
Using the results found at the end of the Eighties \cite{W1, W2, W3, CSFT, Bis1, Bis2} 
and several more recent ideas \cite{collective, NL5}, the string field theory 
has the following schematic structure for the spacetime bosonic and fermionic fields, % \cite{collective, NL5},
\be
&& \hspace{-0.5cm} 
S= \int d^D x \left(  \frac{1}{2} \phi_i K_{i j} (\Box) \phi_j   \label{ESFT} %\\
%&& \hspace{-0.2cm} 
- v_{i j k} \tilde{\phi}_i \tilde{\phi}_j \tilde{\phi}_k \right)  ,
%(e^{\alpha^{\prime} \frac{\ln (3 \frac{\sqrt{3}}{4})}{2} \, \Box} \phi_i)
%(e^{\alpha^{\prime} \frac{\ln (3 \frac{\sqrt{3}}{4})}{2} \, \Box} \phi_j) 
%(e^{\alpha^{\prime} \frac{\ln (3 \frac{\sqrt{3}}{4})}{2} \, \Box} \phi_k)
 %\Big) \,  , \nonumber 
\ee
where 
$$\tilde{\phi}_i \equiv e^{\alpha^{\prime} \frac{\ln (3 \frac{\sqrt{3}}{4})}{2} \, \Box} \phi_i \, , $$
$K_{i j}(\Box) \approx \Box$ for open as well as close bosonic strings, 
and $\alpha^{\prime}$ is the inverse mass square in string theory.
By a field redefinition %and omitting the tensorial structure in the kinetic term 
\cite{collective}, the action (\ref{ESFT}) simplifies to
\be
&& \hspace{-1.2cm} 
S= \! \int \! d^D x \Big(  \frac{1}{2} \phi_i \, \Box \, e^{- \alpha^{\prime} \ln (3 \frac{\sqrt{3}}{4}) \, \Box} \, \phi_j   
%&& \hspace{-0.2cm} 
- v_{i j k} 
 \phi_i \, \phi_j \,   \phi_k
 \Big) . \label{ESFT2} 
\ee
We can immediately observe that
the kinetic term in (\ref{ESFT2}) has the same scaling of the linearized theory studied in this paper for 
the exponential form factor $V_2(z)=\exp( - z)$ ($n=1$). 
If we expand (\ref{actionF}) in powers of the graviton field %without to take cure of 
neglecting the 
exponential factor in the interaction, 
the three-graviton vertex is quite similar to the one in 
(\ref{ESFT2}). However, the general covariance in (\ref{actionF}) implies the same leading scaling in the
kinetic term as well as in the interaction vertexes and we are unable to get a finite theory at any order in the
loop expansion.
As already pointed out in this section, 
one possible loophole to this puzzle could be a supersymmetric extension of the
action in (\ref{action}).
%On the other hand and the comparison is only partial at this stage. 

About the finiteness of string theory, we are likely to endorse the following ideas. 
Due to the presence of the exponential factor, the effective string theory in (\ref{ESFT2}) manifests an asymmetry between the kinetic and the interaction terms.
Contrary to our covariant action (\ref{actionF}), 
such asymmetrical state implies that 
the string theory does not manifest any divergence. %is finite 
%contrary to our covariant action (\ref{actionF}) which anyhow manifests divergences at one loop.
The well-known ``softness" of the high energy tree-level amplitudes also descends from the same asymmetry.  
%between the fields in the quadratic and in the interaction terms. 

However, the comparison here proposed can only be qualitative and partial because, unlike 
the effective string field theory, ours 
is a general covariant theory. Indeed,  
general coordinate invariance in string theory can
 only be achieved through cancellations among contributions from 
infinitely many interactions terms \cite{CSFT}.
%and the comparing can be only qualitative and partial.
However, we do not exclude that a supersymmetric extension of our theory (\ref{actionF}) can 
be framed within ``M-theory" as one of its possible vacuums.

\section*{APPENDIX:
3D HIGHER-DERIVATIVE QUANTUM GRAVITY}
In this section, as a particular toy model, we consider a nonlocal generalization 
of the $3D$ higher derivative gravity studied in \cite{3d0}. % as an example.
The nonlocal action is 
%The theory we are going to introduce is
%a ``nonlocal" extension of the higher derivative gravity in $3D$ 
%and has the following general structure,  
\be \hspace{-0.4cm} 
S = \! \frac{1}{\kappa^2} \! \int \!\! d^3x\sqrt{|g|}   %^4x \sqrt{|g|} 
\left[ \sigma R+ R \, \alpha(\Box) R + R_{\mu \nu}  \beta( \Box)  R^{\mu \nu} \right],
\label{GA}
\ee
% + O(R^3)] \, ,$$
where the two ``form factors" $\alpha(\Box)$ and $\beta(\Box)$ are ``entire functions" of the covariant D'Alembertian operator. %, $\Box_{\Lambda} = \Box/\Lambda^2$ and $\Lambda$ is an invariant mass scale.
%The non locality only involve positive powers of the D'Alembertian covariant operator since the two form 
%factors are entire functions. %, $\Box$. 
We introduce the following definitions,
\be
\hspace{-0.4cm}
 \alpha(\Box)  := \alpha_0 + h_0( - \Box_{\Lambda} ) \,\, , \,\,\, %\nonumber \\
 \beta(\Box) : = \beta_0 + h_2( - \Box_{\Lambda}) \,  ,  
%&& \Box_{\Lambda} : =\Box/\Lambda^2.
\ee
where $\Box_{\Lambda} : =\Box/\Lambda^2$ and $h_0, h_2$ are entire functions.
The two %entire functions 
form factors 
have dimensions:
$[\alpha(\Box)] = [\beta(\Box)] = L^2$.
The Lagrangian, complete with the gauge fixing and ghost terms, reads 
\be
\mathcal{L} = \mathcal{L}_g +  \mathcal{L}_{\rm GF} + \mathcal{L}_{\rm GH}, 
\label{LT}
\ee
where $\mathcal{L}_g$ is expressed by (\ref{GA}), and the graviton fluctuation $h^{\mu \nu}$ is defined by
\begin{eqnarray}
\tilde{g}^{\mu\nu} := \sqrt{ - g} g^{\mu \nu} = \eta^{\mu \nu} + \kappa h^{\mu \nu}. %= \eta_{\mu \nu} +  \kappa h^{\mu \nu} 
\label{graviton}
\end{eqnarray}
Imposing the BRST invariance on the full Lagrangian ({\ref{LT}), we can get the gauge-fixing and 
ghost terms of the action. 
The BRST transformation for the fields in (\ref{LT}) appears 
\be
&& \delta_{\rm B} g_{\mu \nu} = - \delta \lambda \,  [ g_{\rho \nu } \partial_{\mu} c^{\rho} + g_{\rho \mu} \partial_{\nu} c^{\rho} + \partial_{\rho}g_{\mu \nu} c^{\rho} ]  \, ,\nonumber \\
&&  \delta_{\rm B} c^{\mu} = - \delta \lambda c^{\rho} \partial_{\rho} c^{\mu} \,  , \nonumber \\
&&  \delta_{\rm B} \bar{c}_{\mu} = i \delta \lambda \, \omega(\Box) B_{\mu},
\ee
where $c^{\mu}, \bar{c}_{\nu}$ are the anti-commuting ghosts fields, $B_{\mu}$ is the auxiliary field, 
$\delta \lambda$ is an anti-commuting constant parameter and $\omega(\Box)$ is a weight function 
of the d'Alembertian operator. % on the flat spacetime. 
The dimensions of the fields are: $[h] = L^{-1/2}$, $[c]= L^y$, $[\bar{c}] = L^{-y-1/2}$, 
$[B]=L^{-3/2}$, 
$[\delta \lambda]= L^{1-y}$.
The BRST transformation for the graviton field in (\ref{graviton}) can be extracted from  %\cite{3d0} %read from
\be
&& 
\hspace{-0.55cm} 
\delta_B \tilde{g}_{\mu \nu} = \delta \lambda ( \tilde{g}^{\mu \rho} \partial_{\rho} c^{\nu} + 
\tilde{g}^{\nu \rho} \partial_{\rho} c^{\nu} 
- \tilde{g}^{\mu \nu} \partial_{\rho} c^{\rho}
- \partial_{\rho} \tilde{g}^{\mu \nu} \,  c^{\rho} \nonumber \\
&& \hspace{0.5cm}\equiv \delta \lambda \, \mathcal{D}^{\mu \nu}_{\rho} \, c^{\rho},
\ee
which implies $\delta_B h^{\mu \nu} = \kappa  \delta \lambda \, \mathcal{D}^{\mu \nu}_{\rho} \, c^{\rho}$ \cite{3d0}. 
%\be
%&& \hspace{-0.3cm}
%\mathcal{L} =   - \sqrt{ - g} \, \Bigg[ \frac{\beta}{\kappa^2} R
%- R_{\mu \nu} \Big( \beta_2 - h_2( - \Box_{\Lambda} \Big) R^{\mu \nu} \label{theory} \\
%&& \hspace{-0.3cm} + R \Bigg(  \frac{\beta_2}{3} + \beta_0   -   h_0 ( - \Box_{\Lambda}) 
%- \frac{h_2( - \Box_{\Lambda}  ) }{3}  \Bigg)   R \Bigg] + \mathcal{L}_{\rm GF, GH} ,
%\nonumber 
%\nonumber 
%&&
% \left(R^{\mu \nu} - \frac{1}{2} g^{\mu \nu} R \right) 
 %G^{\mu \nu} F(\Box/\Lambda^2) R_{\mu \nu},
%\ee
The gauge fixing and ghost Lagrangian can both 
be expressed as a BRST variation of the following functional 
\be
 && 
 \hspace{-0.5cm} \mathcal{L}_{\rm GF} + \mathcal{L}_{GH}= 
 i \delta_B \left[ \bar{c}_{\mu} ( \partial_{\nu} h^{\mu \nu} - a \, B^{\mu}/2 )\right] \frac{1}{\delta \lambda} 
\label{GFG} \\
 && \hspace{-0.5cm}  
 = 
 - B_{\mu} \omega(\Box) \partial_{\nu} h^{\mu\nu} - i \kappa \bar{c}_{\mu} \mathcal{D}^{\mu \nu}_{\rho} c^{\rho} +
 \frac{a}{2} B_{\mu} \omega(\Box) B^{\mu}.
\nonumber 
\ee
To obtain the graviton propagator, we first eliminate the auxiliary field $B_{\mu}$ to get the following 
gauge fixing Lagrangian, 
\be
 \mathcal{L}_{\rm GF} = %+ \mathcal{L}_{GH}= 
 - \frac{1}{2 a} \, (\partial_{\nu} h^{\mu \nu}) \, \omega(\Box) \, (\partial_{\nu} h^{\mu \nu})
 %- i \kappa \bar{c}_{\mu} \mathcal{D}^{\mu \nu}_{\rho} c^{\rho}
\label{NoB}
\ee
and then we assemble the
quadratic part of (\ref{LT}), namely %which reads 
\begin{eqnarray}
&& \hspace{-0.4cm} 
\mathcal{L} = \frac{1}{4} h^{\mu \nu} \, \Box \, [ 
P^{(2)} ( \Box \, \Box \beta(\Box) + \sigma)  \label{Q3d} \\
&& \hspace{-0.4cm} 
+ \Box \, \omega(\Box) \, P^{(1)} /a 
+ P^{(0,s)} ( (8 \, \alpha(\Box) + 3 \, \beta(\Box) ) - \sigma) \nonumber \\
&& \hspace{-0.4cm} 
+ 2 P^{(0, \omega )} ( (8 \, \alpha(\Box) + 3 \, \beta(\Box) )\Box - \sigma + \omega(\Box)/a) + \sqrt{2} 
\times 
\nonumber \\
&& \hspace{-0.4cm} 
( P^{(0, s \omega )} + P^{(0, \omega s)}) ( ( 8 \, \alpha(\Box) + 3 \, \beta(\Box) )\Box - \sigma) ]_{\mu \nu, \rho \sigma}
h^{\rho \sigma}.
\nonumber 
 \end{eqnarray}
In (\ref{Q3d}) we have introduced the $3D$ projectors. % following $3D$ 
%projectors %$P^{(2)}, P^{(1)},  P^{(0,s)}, P^{(0,s\omega)}, P^{(0,\omega s)} $ 
%\cite{VN}, 
%{\small 
 %
 %
 %\hspace{-0.299cm}
Using the orthogonality and the completeness property of the projectors, %(\ref{proje2})
we find the 
graviton propagator 
{\small 
\be
&& 
\hspace{-0.4cm} 
D(k) = \frac{1}{(2 \pi)^3 \, k^2} \Bigg[ \frac{P^{(2)}}{\beta(k^2) k^2 - \sigma} + 
\frac{P^{(0,s)}}{ (8 \alpha(k^2) + 3 \beta(k^2)) k^2 + \sigma} \nonumber \\
&& \hspace{-0.5cm}
- \frac{a}{ \omega(k^2)} \Big(  P^{(1)} +  P^{(0,s)} + \frac{P^{(0,\omega)}}{2} %\nonumber \\
%&& \hspace{-0.5cm}
- \frac{\sqrt{2}}{2}  ( P^{(0,s \omega)} + P^{(0,\omega s)} ) \Big)\Bigg] . %\frac{1}{(2 \pi)^3}.
\nonumber %\label{propa3D}
\ee}
\hspace{-0.25cm}
In the harmonic gauge $\partial_{\mu} h^{\mu \nu} =0$ (or $a=0$), the propagator 
considerably simplifies to
\be
D(k) = \frac{1}{(2 \pi)^3k^2} \Bigg[ \frac{P^{(2)}}{\bar{h}_2 \left(\frac{k^2}{\Lambda^2} \right)} + 
\frac{P^{(0,s)}}{ \bar{h}_0 \left(\frac{k^2}{\Lambda^2} \right)} \Bigg] \, , 
\label{propharmonic}
\ee
where the following notation has been introduced,
\be
&& \bar{h}_2(z) := \beta(z) z \Lambda^2 - \sigma \, , \nonumber \\
%\beta - \beta_2 \kappa^2 \Lambda^2 z + \kappa^2 \Lambda^2 z \, h_2(z) , \nonumber \\
&& \bar{h}_0(z) := (8 \alpha(z) + 3 \beta(z)) z \Lambda^2 + \sigma, \nonumber \\
&& z := - \Box_{\Lambda}.
\label{hbar}
\ee 
%and $z := - \Box_{\Lambda}$, 
As in the $D$-dimensional case (\ref{hz}), we choose the entire functions $h_2(z), h_0(z)$
compatibly with the properties (i)-(iii),
 \be
&& h_2(z) = - \frac{\sigma (V(z)^{-1} -1) + \tilde{\beta_0} \Lambda^2 z}{\Lambda^2 z} \, ,\nonumber \\
&& h_0(z) = \frac{4 \sigma (V(z)^{-1} -1) - 8 \tilde{\alpha_0} \Lambda^2 z}{8 \Lambda^2 z}.
\ee
Assuming the theory to be renormalized at a particular scale $\mu_0$, we identify
\be
\tilde{\alpha}_0 = \alpha_0(\mu_0) \,\, , \,\,\,\,\, \tilde{\beta}_0 = \beta_0(\mu_0).
\ee
In this case $\bar{h}_2 = \bar{h}_0 = V(z)^{-1}$ and the propagator simplifies to
\be
D(k) = - \frac{V\left( \frac{k^2}{\Lambda^2} \right)}{(2 \pi)^3 \, \sigma \, k^2} [ P^{(2)} - P^{(0,s)}  ] \, .
\label{propharmonic2}
\ee
%Sihe form factor $V(z)$, % extensively explained in the paper 
The pole structure of the propagator is the same one as that in the local theory,
because $V(z)$ is an entire function with no zeros in the complex plane. 
In $D=3$ there are no local degrees of freedom, and therefore the
amplitude (\ref{ampli2}) is identical to zero. 

What we have presented in this section is a toy model of modified nonlocal gravity 
that 
we think it might be interesting to expand on in connection with
%the same analysis can be extended to 
other 
three-dimensional theories studied in recent years 
\cite{3d1, 3d2, 3d3, 3d4, 3d5, 3d6, 3d7, 3d8, 3d9, 3d10, 3d11, 3d12}.

\begin{acknowledgments}
\noindent %P.N. is supported by the Helmholtz International Center for FAIR within the
%framework of the LOEWE program (Landesoffensive zur Entwicklung Wissenschaftlich-\"{O}konomischer Exzellenz) launched by the State of Hesse. P.N. would like to thank the Perimeter Institute for Theoretical Physics, Waterloo, ON, Canada for the kind hospitality during the period of work on this project. 
A special thanks goes to Yuri Gusev for drawing our attention to a large body of literature on 
nonlocal field theory. 
We also thank Eugenio Bianchi, Dario Benedetti Tirthabir Biswas, Gianluca Calcagni, Francesco Caravelli, Enore Guadagnini, Anupam Mazumbar, John Moffat, %Alberto Montina, 
Tim Koslowski, Gabor Kunstatter, Roberto Percacci, Jorge Russo, Ilya Shapiro, Lee Smolin  %Gabor Kunstatter 
and Pasquale Sodano.
Research at Perimeter Institute is
supported by the Government of Canada through Industry 
Canada and by the Province of Ontario through the
Ministry of Research \& Innovation.
\end{acknowledgments}

%\newpage

\end{document}